\documentclass[11pt]{article}
\usepackage{amssymb,amsmath,epsfig}

\begin{document}

\title{\bf Reconstructions of $f(T)$ Gravity from Entropy Corrected
Holographic and New Agegraphic Dark Energy Models in Power-law and
Logarithmic Versions}

\author{\bf{Pameli Saha}\thanks{pameli.saha15@gmail.com}~ and
\bf{Ujjal Debnath}\thanks{ujjaldebnath@gmail.com}\\
Department of Mathematics, Indian Institute of Engineering\\
Science and Technology, Shibpur, Howrah-711 103, India.\\}

\date{}
\maketitle

\begin{abstract}
Here, we peruse cosmological usage of the most promising
candidates of dark energy in the framework of $f(T)$ gravity
theory where $T$ represents the torsion scalar teleparallel
gravity. We reconstruct the different $f(T)$ modified gravity
models in the spatially flat Friedmann-Robertson-Walker (FRW)
universe according to entropy-corrected versions of the
holographic and new agegraphic dark energy models in power-law and
logarithmic corrections, which describe accelerated expansion
history of the universe. We conclude that the equation of state
parameter of the entropy-corrected models can transit from
quintessence state to phantom regime as indicated by recent
observations or can lie entirely in the phantom region. Also,
using these models, we investigate the different erase of the
stability with the help of the squared speed of sound.
\end{abstract}


\section{Introduction}

The type Ia Supernovae and Cosmic Microwave Background (CMB)
\cite{Perlmutter,Riess} observations point out that our universe
is precisely accelerating which is caused by some unknown fluid
having positive energy density and negative pressure, called as
``Dark Energy" (DE). Observations indicate that dark energy
occupies about 70\% of the total energy of the universe, whereas
the contribution of dark matter is 26\% and rest 4\% is the
baryonic matter. For related review works see the references
\cite{Copeland,Tsujikawa}. Although a long-time argument has been
made on this interesting issue of modern cosmology, we still have
a few knowledge about DE. The cosmological constant $\Lambda$ is
the most appealing and simplest candidate for DE which obeys the
equation of state parameter $w=-1$. However, the cosmological
constant suffers from two serious theoretical problems, i.e., the
cosmological constant problem and the coincidence problem. In this
respect, different dynamical DE models and different modified
theories of gravity have been developed. Moreover, the
reconstruction phenomenon of different DE models
\cite{Sahni,Seikel,Clarkson,Liu} gains great
attention to discuss the accelerated expansion of the universe.\\

In recent years, an interest has been proposed to study the dark
energy in the new form i.e., Holographic Dark Energy (HDE) model
\cite{Enqvist1,Zhang1,Pavon} which arises from the holographic
principle \cite{Fischler} stating that the number of degrees of
freedom of a physical model must be finite \cite{Hooft} and an
infrared cut-off should constrain it \cite{Cohen}. In quantum
field theory \cite{Cohen}, for developing a black hole , the UV
cut-off $\Lambda$ should relate with the IR cut off $L$ due to
limit set. In the reference \cite{Li1} by Li, he debated a
relation $L^{3} \rho_{\Lambda}\leq L M_{P}^{2}$, where
$\rho_{\Lambda}$ is the quantum zero point energy density and
$M_{P}=\frac{1}{\sqrt{8\pi G}}$ is the reduced Plank Mass i.e.,
the mass of a black hole of the size $L$ should not be exceeded by
the total energy in a region of same size. The HDE models have
been discussed in \cite{Enqvist,Huang0,Gong,Elizalde,Zhang0}. The
black hole entropy $S_{BH}$ plays an important role in the
simplification of HDE, given as usually, $S_{BH}=\frac{A}{4G}$,
where $A\sim L^{2}$ is the area of horizon.\\

The power-law corrections arise in dealing with the entanglement
of quantum fields moving into and out of the horizon
\cite{Radicella,Sheykhi,Das} for which the entropy-area relation
for power-law correction can be given as

\begin{equation}\label{1}
S_{BH}=\frac{A}{4G}[1-K_{\epsilon}A^{1-\frac{\epsilon}{2}}],
\end{equation}
where
\begin{equation*}
K_{\epsilon}=\frac{\epsilon
(4\pi)^{\frac{\epsilon}{2}-1}}{(4-\epsilon)r_{c}^{2-\epsilon}}
\end{equation*}

Here, $r_{c}$ is the crossover scale and $\epsilon$ is the
dimensionless constant. Motivated by this corrected entropy-area
relation (1) in the setup of LQG (loop quantum gravity), Wei
\cite{Wei} suggested the energy density of the ECHDE in Power-law
Correction.\\

Also the of entropy-area relation for a logarithmic correction can
be improved to \cite{Banerjee1,Modak,Sadjadi1,Wei0}

\begin{equation}\label{2}
S_{BH}=\frac{A}{4G}+\alpha \ln[\frac{A}{4G}]+\beta,
\end{equation}\\

where $\alpha$ and $\beta$ are dimensionless constants of order
unity. Recently, inspired by the corrected entropy-area relation
(2) in the setup of LQG, Wei \cite{Wei} propounded the energy
density of the entropy-corrected HDE (ECHDE)in Logarithmic
Correction.\\

From quantum mechanics along with the gravitational purpose in
General Relativity, the another type of dark energy is the
agegraphic DE (ADE) model. The original agegraphic DE model was
brought by Cai \cite{Cai1} to study the accelerating expansion of
the universe where the age ($T$) of the universe is present in the
expression of energy density, given by
\begin{equation}\label{3}
\rho_{\Lambda}=3c^{2}M_{P}^{2}T^{-2}
\end{equation}
The numerical factor $3c^{2}$ is used to recover some
uncertainties. Subsequently, Wei and Cai \cite{Wei and Cai}
suggested a new kind of ADE model by removing the age of the
universe and placing with the conformal time ($\eta$), called as
new agegraphic DE (NADE) model. Recently, Wei \cite{Wei} initiated
the energy density of the entropy-corrected NADE (ECNADE) in
power-law and logarithmic corrections like the entropy-corrected
HDE (ECHDE) in power law and logarithmic corrections model and
details of these were discussed in
\cite{Karami01,Karami02,Karami03,Farooq1,Malekjani}.\\

There is an another discussion for the cosmic acceleration of the
Universe (predict from observational data) , so-called ``modified
gravity" where we do not require any additional components like DE
(for review see \cite{Capozziello}) for acceleration of the
Universe. Various kinds of modified theories have been proposed
such as $f(R)$ \cite{Nojiri0}, $f(G)$ \cite{Myrzakulov,
Banijamali}, Horava-Lifshitz \cite{Kiritsis} and Gauss-Bonnet
\cite{Nojiri and Odintsov; Li} theories of gravity. Recently,
\cite{Cai and Chen, Ferraro} formulate a new kind of theory of
gravity known as $f(T)$ gravity in a space-time possessing
absolute parallelism. $f(T)$ gravity have been recently studied in
\cite{Li00,Sotiriou}. In the $f(T)$ theory of gravity, the
teleparallel Lagrangian density gave a description of the torsion
scalar $T$, evoked to be a function of $T$, i.e., $f(T)$, for the
late time cosmic acceleration \cite{Bamba}. In a recent work,
Jamil et al \cite{Jamil and Yesmakhanova,Jamil and Momeni}
investigated the interacting DE model and state-finder diagnostic
in $f(T)$
cosmology.\\

Recently, the reconstruction of various types of modified
gravities $f(R)$, $f(T)$, $f(G)$, Einstein-Aether etc. with the
various dark energy models have made a plea topic in cosmology
\cite{Khodam,Jawad,Chattopadhyay,Debnath1,Hamani}. Farooq et al
\cite{Farooq} reconstructed $f(T)$ and $f(R)$ gravity according to
$(m,n)$-type Holographic dark energy, Karami et al. \cite{Karami}
did the reconstruction of $f(R)$ modified gravity from ordinary
and entropy-corrected versions of holographic and new agegraphic
dark energy models and also Debnath \cite{Debnath0} discussed on
the topic of reconstruction of $f(R)$, $f(G)$, $f(T)$ and
Einstein-Aether gravities from entropy-corrected $(m,n)$ type
pilgrim dark energy. Motivated by these works, with the help of
the modified $f(T)$ gravity and considering the entropy-corrected
versions of the HDE and NADE scenarios, it is interesting to
investigate how the $f(T)$-gravity can describe ECHDE and ECNADE
densities in power-law and logarithmic versions as effective
theories of DE models. This paper is arranged as follows. In
section 2, we give a brief idea of the theory of $f(T)$ gravity
and corresponding solutions for FRW background. In sections 3 and
4, we reconstruct the different $f(T)$ gravity models i.e., find
unknown function $f(T)$ corresponding to the ECHDE and ECNADE
models in power-law and logarithmic versions, respectively and
analyze the EoS parameter for the corresponding models. Karami et
al \cite{K} also investigated the modified teleparallel gravity
models as an alternative for holographic and new agegraphic dark
energy models. In section 5, we provide the analysis and
comparison of the reconstructed
models. Section 6 is invoked to our conclusions.\\

\section{The brief idea of $f(T)$ gravity and ECHDE in power-law and logarithmic correction:}

Teleparallel gravity is correlated with a gauge theory for the
translation group. For unusual character of this translations, any
gauge theory with these translations is different from the usual
gauge theory in many ways, mostly in the background of tetrad
field whereas this field is used to define a linear Weitzenbock
connection, presenting torsion without no curvature. For the
details of this gravity theory see the review \cite{Andrade}. We
consider here to generalize the teleparallel Lagrangian $T$ to a
function $f(T)=T+g(T)$, which is same as the generalization of the
Ricci scalar in Einstein-Hilbert action to the modified $f(R)$
gravity. We can write the action of $f(T)$ gravity, coupled with
matter $L_{m}$ by \cite{Bengochca, Linder0,Cai000,Setare00,Setare
and Houndjo}
\begin{equation}\label{4}
S=\frac{1}{16\pi G}\int d^{4}x e(T+g(T)+L_{m})
\end{equation}
where $e=det(e^{i}_{\mu})=\sqrt{-g}$. Now we will take the units
$8\pi G=c=1$. Here, the teleparallel Lagrangian $T$, known as the
torsion scalar, is defined as follows:
\begin{equation}\label{5}
T=S_{\rho}^{\mu\nu}T_{\mu\nu}^{\rho},
\end{equation}
where
\begin{equation}\label{6}
T_{\mu\nu}^{\rho}=e_{i}^{\rho}\left(\partial_{\mu}e_{\nu}^{i}-\partial_{\nu}e_{\mu}^{i}\right),
\end {equation}
\begin{equation}\label{7}
S_{\rho}^{\mu\nu}=\frac{1}{2}(K^{\mu\nu}_{\rho}+\delta^{\mu}_{\rho}T^{\theta\nu}_{\theta}
-\delta^{\nu}_{\rho}T^{\theta\mu}_{\theta}),
\end{equation}
and $K^{\mu\nu}_{\rho}$ is the contorsion tensor
\begin{equation}\label{8}
K^{\mu\nu}_{\rho}=-\frac{1}{2}(T^{\mu\nu}_{\rho}-T^{\nu\mu}_{\rho}-T^{\mu\nu}_{\rho}),
\end{equation}
Making a variation of the action with respect to vierbein
$e^{i}_{\mu}$, we get the field equations as
\begin{equation}\label{9}
e^{-1}\partial_{\mu}(eS_{i}^{\mu\nu}\left)(1+g_{T}\right)-e_{i}^{\lambda}
T_{\mu\lambda}^{\rho}S_{\rho}^{\nu\mu}g_{T}+S_{i}^{\mu\nu}\partial_{\mu}(T)g_{TT}-\frac{1}{4}
e_{i}^{\nu}(1+g(T))=\frac{1}{2} e_{i}^{\rho}\Upsilon_{\rho}^{\nu},
\end{equation}
where $g_{T}$ and $g_{TT}$ are the first and second derivatives of
$g$ with respect to $T$. Here $\Upsilon_{\rho\nu}$ is the stress
tensor. Now we assume the usual spatially flat metric of the
Friedmann-Robertson-Walker (FRW) universe giving the line element
written as
\begin{equation}\label{10}
ds^{2}=dt^{2}-a^{2}(t)\sum_{i=1}^{3}(dx^{i})^{2}
\end{equation}
where $a(t)$ is the scalar factor , a function of the cosmic time
$t$. Moreover, we consider the background to be a perfect fluid.
Using the FRW metric and the perfect fluid matter in the
Lagrangian (5) and the field equation (9), we obtain
\begin{equation}\label{11}
T=-6H^{2},
\end{equation}
\begin{equation}\label{12}
3H^{2}=\rho-\frac{1}{2}g-6H^{2}g_{T},
\end{equation}
\begin{equation}\label{13}
-3H^{2}-2\dot{H}=p+\frac{1}{2}g
+2(3H^{2}+\dot{H})g_{T}-24\dot{H}H^{2}g_{TT},
\end{equation}
where $\rho$ and $p$ are the energy density and pressure of
ordinary matter content of the universe, respectively. The Hubble
parameter ($H$) is defined as $H=\frac{\dot{a}}{a}$, where the
``dot" denotes the derivative with respect to the cosmic
time. Equation (11) shows that $T<0$.\\

The equation of state (EoS) parameter due to the torsion
contribution is defined as
\begin{equation}\label{14}
w_{\Lambda}=\frac{p_{\Lambda}}{\rho_{\Lambda}}
\end{equation}
which shows that for the phantom, $w_{\Lambda}<-1$, and
quintessence,
$w_{\Lambda}>-1$, dominated universe.\\

We define the redshift $z$ as

\begin{equation*}
1+z=\frac{a_{0}(t)}{a(t)}
\end{equation*}
where $a_{0}(t)$=1 for the present epoch.\\

For a given $a(t)$, by the help of equations (12) and (13) one can
reconstruct the $f(T)$ gravity according to any DE model given by
the EoS $p_{\Lambda}=p_{\Lambda}(\rho_{\Lambda})$ i.e.,
$\rho_{\Lambda}=\rho_{\Lambda}(a)$. There are two classes of scale
factors which usually people consider them for describing the
accelerating universe in
$f(R)$, $f(T)$, etc.\\

{\bf Class I:} The first class of scale factor is given by
\cite{Setare},
\begin{equation}\label{15}
a(t)=a_{0}(t_{s}-t)^{-n},~t\leq t_{s}
\end{equation}
where $a_{0}$, $n$ are constants and $t_{s}$ defines the future
singularity time. Hence,
\begin{equation}\label{16}
H=\frac{n}{t_{s}-t}
\end{equation}
\begin{equation}\label{17}
T=-\frac{6n^{2}}{(t_{s}-t)^{2}}
\end{equation}

{\bf Class II:} For the second class of scale factor defined as
\cite{Setare},
\begin{equation}\label{18}
a(t)=a_{0}t^{n},n>0
\end{equation}
One can obtain,
\begin{equation}\label{19}
H=\frac{n}{t}
\end{equation}
\begin{equation}\label{20}
T=-\frac{6n^{2}}{t^{2}}
\end{equation}\\

For the both cases we get
\begin{equation}\label{21}
z=(n\sqrt{\frac{6}{-T}})^{n}-1
\end{equation}\\

Using the two classes of scale factors (15) and (18), we
reconstruct the different $f(T)$ gravities according
to the ECHDE and ECNADE models in power-law and logarithmic versions.\\

\section{$f(T)$ reconstruction from ECHDE in power-law and logarithmic corrections models}

\subsection{ECHDE in power-law correction}

\cite{Wei} proposed the energy density of the ECHDE in power-law
correction using the relation (1) as \cite{Wei}

\begin{equation}\label{22}
\rho_{\Lambda}=3\delta^{2}R_{h}^{-2}-\lambda R_{h}^{-\epsilon}
\end{equation}\\

where $\lambda$ is a constant related with $\epsilon$ and
$K_{\epsilon}$, $\delta$ is a constant.In the special case
$\lambda=0$, the above equation reduces to the well-known HDE
density. Also $R_{h}$ is the future event horizon defined as

\begin{equation}\label{23}
R_{h}=a\int_{t}^{t_{s}}\frac{dt}{a}
\end{equation}\\

For the first class (class I) of scale factor (15) and using
equation (16), the future event horizon $R_{h}$ yields
\begin{equation}\label{24}
R_{h}=a(t)\int_{t}^{t_{s}}\frac{dt}{a(t)}=\frac{t_{s}-t}{n+1}=\sqrt{-\frac{6n^{2}}{T(n+1)^{2}}}
\end{equation}
Replacing equation (24) into (22) one can get
\begin{equation}\label{25}
\rho_{\Lambda}=\frac{\delta^{2}(n+1)^{2}(-T)}{2n^{2}}-\lambda
(\frac{n+1}{\sqrt{6}n})^{\epsilon}(-T)^{\frac{\epsilon}{2}}
\end{equation}\\

Substituting equation (25) in the differential equation (12) i.e.,
$\rho=\rho_{\Lambda}$, gives the following solution

\begin{equation}\label{26}
f(T)=c\sqrt{-T}-\frac{\delta^{2}(n+1)^{2}(-T)}{n^{2}}+\frac{2\lambda}{\epsilon
-1)}(\frac{n+1}{\sqrt{6}n})^{\epsilon}(-T)^{\frac{\epsilon}{2}}
\end{equation}

where $c$ is the integration constant to be determined from the
necessary boundary condition. In figure {\bf 1}, we understand
that $f(T)\nrightarrow 0$ as $T\rightarrow 0$ for the solution
obtained from equation (26). We also observe that $f(T)$ first
decreases and then increases as $T$ increases keeping in the mind
that $f(T)$ takes always negative value for all values of negative
$T$. Replacing equation (26) into (13) and using (25) we obtain
the EoS parameter of the ECHDE $f(T)$ gravity in power-law
correction model as
$w_{\Lambda}=\frac{p_{\Lambda}}{\rho_{\Lambda}}$ graphically. In
figures {\bf 3} and {\bf 5}, we see that the EoS parameter can
justify the transition from quintessence state $w_{\Lambda}>-1$,
to the phantom regime, $w_{\Lambda}<-1$, i.e., it crosses the
phantom divide line $w_{\Lambda}=-1$ if we draw the graph of EoS
parameter with $T$ and $z$ using the equation (21) respectively.
So in this case, $f(T)$ gravity
generates phantom crossing.\\

For the second class (class II) of scale factor (18) and using
equation (19), the future event horizon $R_{h}$ yields
\begin{equation}\label{27}
R_{h}=a(t)\int_{t}^{\infty}\frac{dt}{a(t)}=\frac{t}{(n-1)}=\sqrt{\frac{-6n^{2}}{T(n-1)^{2}}}
\end{equation}
Replacing equation (27) into (22) one can get
\begin{equation}\label{28}
\rho_{\Lambda}=\frac{\delta^{2}(n-1)^{2}(-T)}{2n^{2}}-\lambda
(\frac{n-1}{\sqrt{6}n})^{\epsilon}(-T)^{\frac{\epsilon}{2}}
\end{equation}
Substituting equation (28) in the differential equation (12) i.e.,
$\rho=\rho_{\Lambda}$, gives the following solution
\begin{equation}\label{29}
f(T)=c\sqrt{-T}-\frac{\delta^{2}(n-1)^{2}(-T)}{n^{2}}+\frac{2\lambda}{\epsilon
-1)}(\frac{n-1}{\sqrt{6}n})^{\epsilon}(-T)^{\frac{\epsilon}{2}}
\end{equation}

where $c$ is the integration constant to be determined from the
necessary boundary condition. In figure {\bf 2}, we understand
that $f(T)\nrightarrow 0$ as $T\rightarrow 0$ for the solution
obtained from equation (29). We also observe that $f(T)$ first
increases and then decreases as $T$ increases keeping in the mind
that $f(T)$ takes always positive value for all values of negative
$T$.It may be stated that the solutions obtained in equations (26)
and (29) are not so-realistic models. Replacing equation (29) into
(13) and using (28) we obtain the EoS parameter of the ECHDE
$f(T)$ gravity in power-law correction model as
$w_{\Lambda}=\frac{p_{\Lambda}}{\rho_{\Lambda}}$ graphically. In
figures {\bf 4} and {\bf 6}, we see that the EoS parameter wholly
lies in the phantom region i.e., $w_{\Lambda}<-1$ always if we
draw the graph of EoS parameter with $T$ and $z$ using the
equation (21) respectively. So in this case, $f(T)$ gravity does
not generate phantom crossing.  \\

\begin{figure}

\includegraphics[height=2.0in]{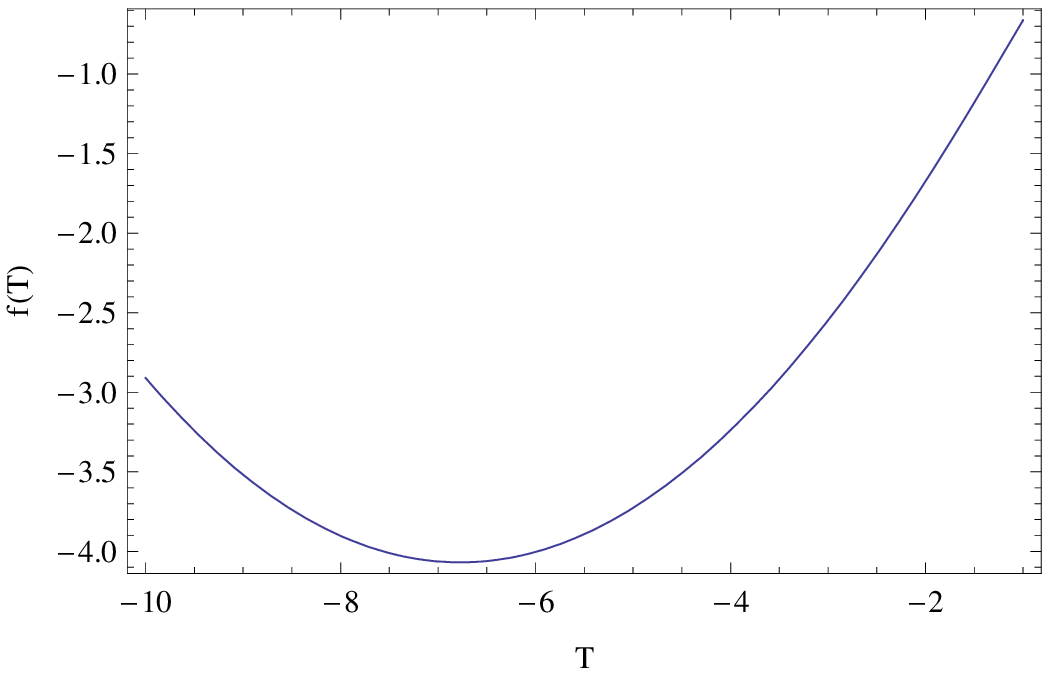}~~~~~
\includegraphics[height=2.0in]{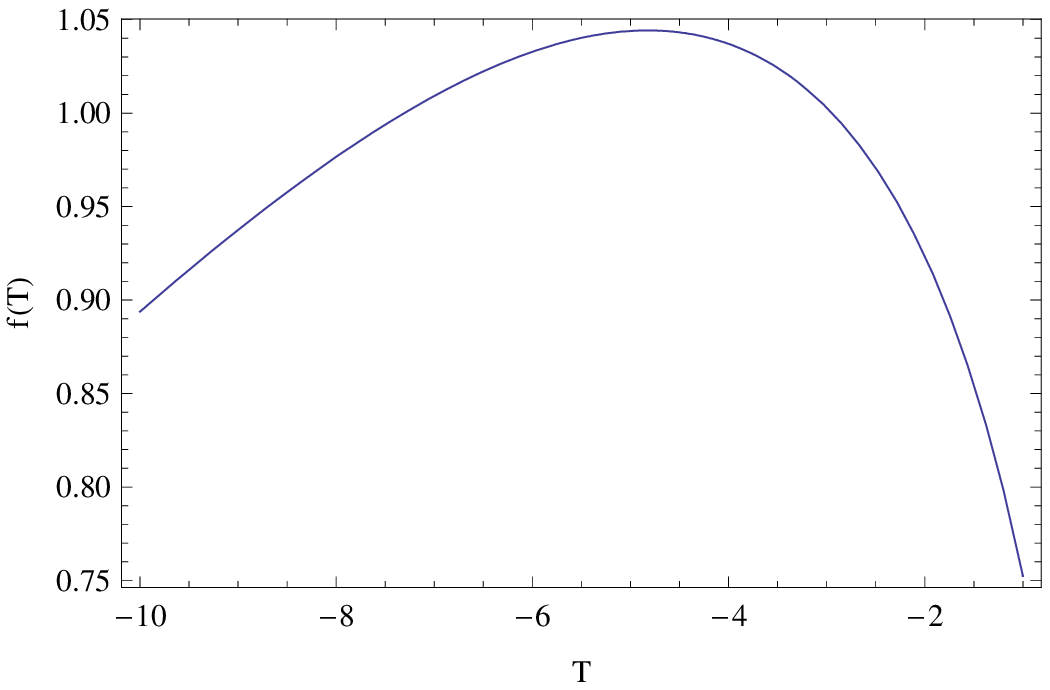}
\vspace{4 mm}
~~~~~~~~~~~~~~~~~~~~~~~Fig.1 ~~~~~~~~~~~~~~~~~~~~~~~~~~~~~~~~~~~~~~~~~~~~~~~~~~~~~~~~~~~~~~~~~~~~~~~Fig.2 \\
\vspace{4 mm}

\includegraphics[height=2.0in]{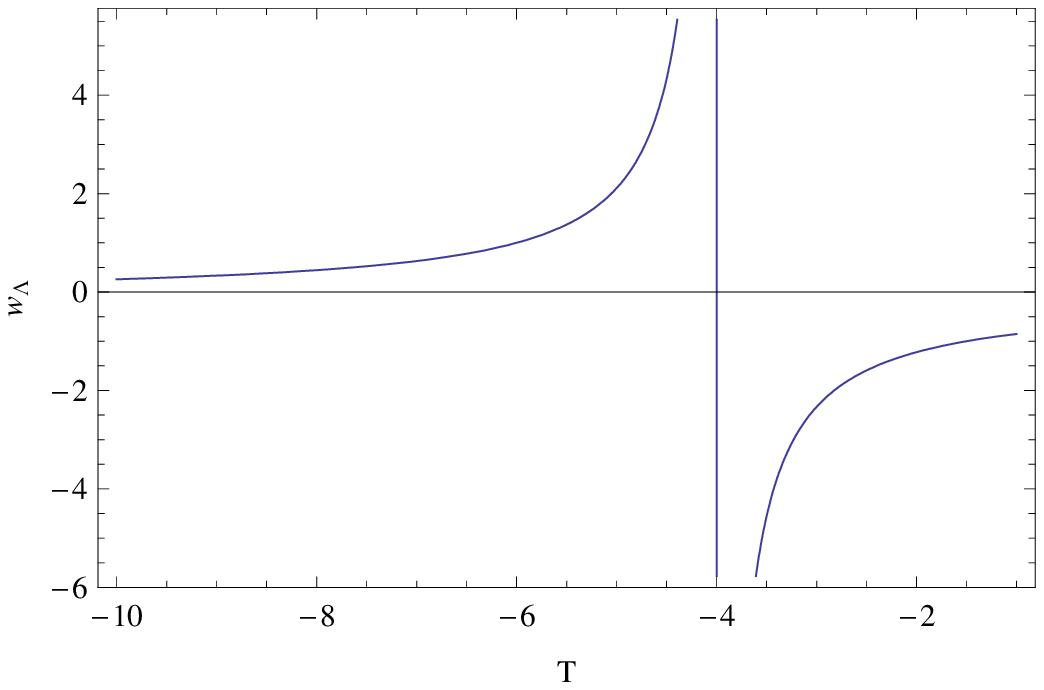}~~~~~
\includegraphics[height=2.0in]{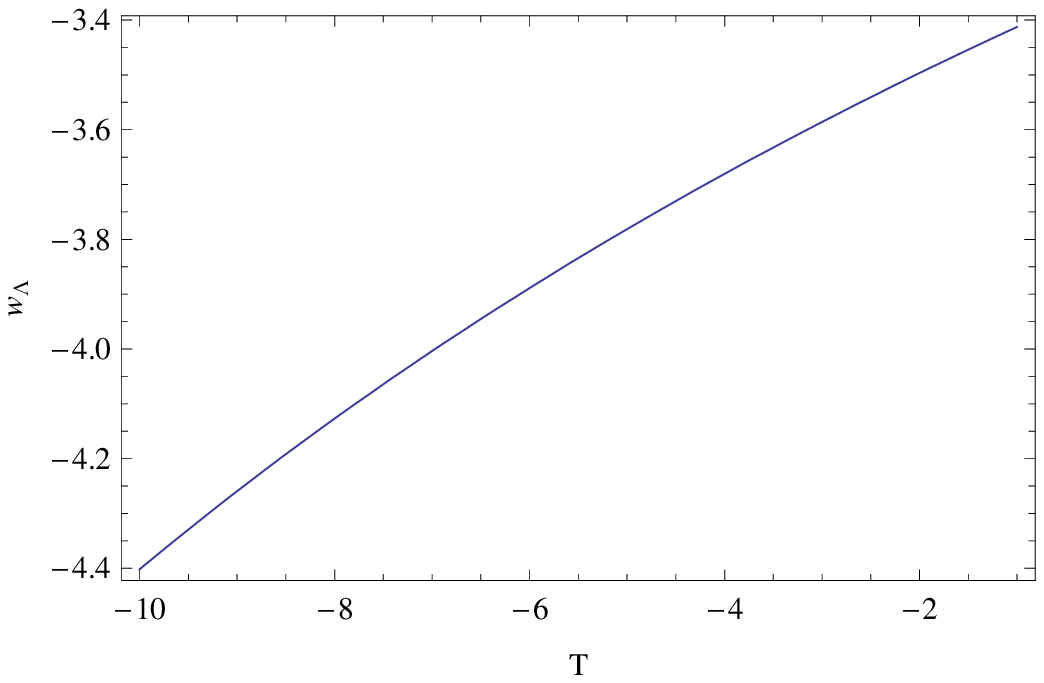}
\vspace{4 mm}
~~~~~~~~~~~~~~~~~~~~~~~Fig.3 ~~~~~~~~~~~~~~~~~~~~~~~~~~~~~~~~~~~~~~~~~~~~~~~~~~~~~~~~~~~~~~~~~~~~~~~Fig.4 \\
\vspace{4 mm}

\includegraphics[height=2.0in]{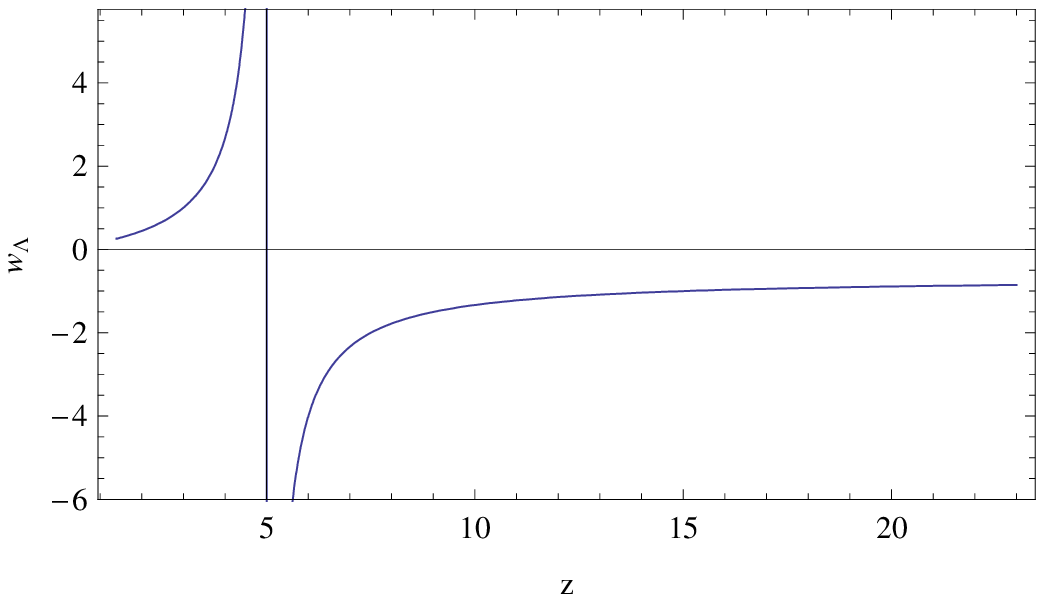}~~~~~
\includegraphics[height=2.0in]{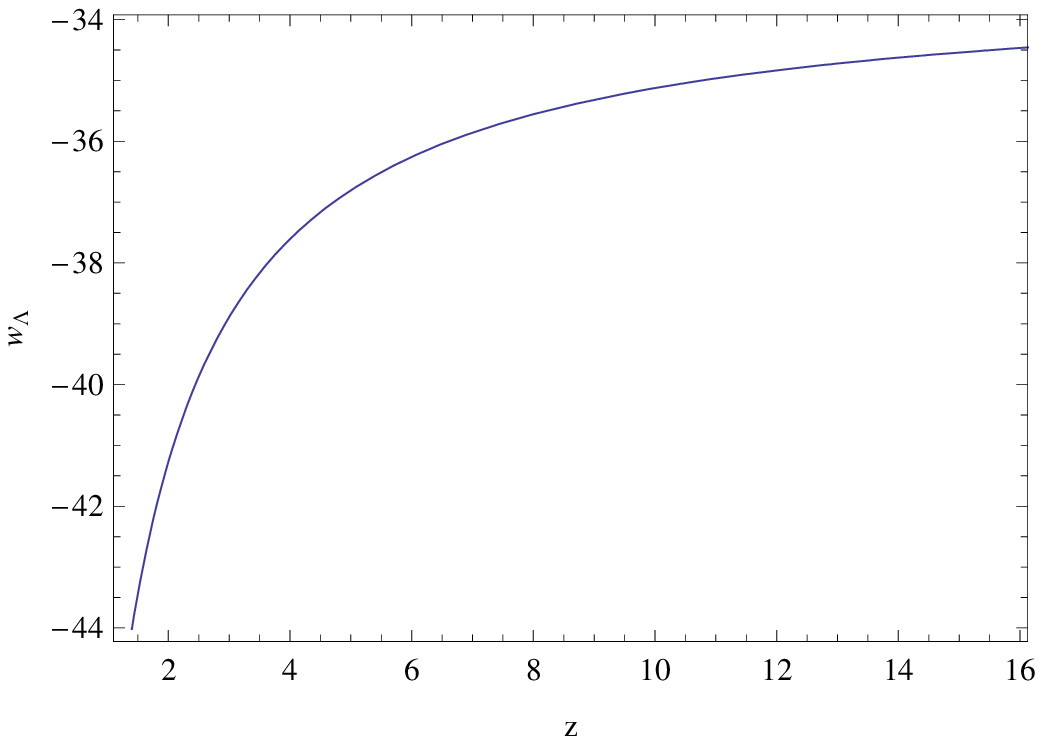}
\vspace{4 mm}
~~~~~~~~~~~~~~~~~~~~~~~Fig.5 ~~~~~~~~~~~~~~~~~~~~~~~~~~~~~~~~~~~~~~~~~~~~~~~~~~~~~~~~~~~~~~~~~~~~~~~Fig.6 \\
\vspace{4 mm}\\
 \textbf{Figs. 1, 3 and 5} represent the plots of $f(T)$ and
$w_{\Lambda}$ for class I scale factor in ECHDE $f(T)$ gravity in
power-law correction model. \textbf{Figs. 2, 4 and 6} represent
the plots of $f(T)$ and $w_{\Lambda}$ for class II scale factor in
ECHDE $f(T)$ gravity in power-law correction model.
\end{figure}

\subsection{ECHDE in logarithmic correction}

\cite{Wei} described the energy density of the ECHDE in
logarithmic version using the corrected entropy-area relation (2)
as \cite{Wei}
\begin{equation}\label{30}
\rho_{\Lambda}=\frac{3\delta^{2}}{R_{h}^{2}}+\frac{\alpha}{R_{h}^{4}}\ln(R_{h}^{2})+\frac{\beta}{R_{h}^{4}}
\end{equation}
where $\alpha$ and $\beta$ are dimensionless constants of order
unity and $\delta$ is a constant.In the special case
$\alpha=\beta=0$, the above equation becomes the well-known HDE
density. Since for only $R_{h}$ being very small, the last two
terms in equation (30) can be comparable to the first term , the
corrections is sensible only at the early stage of the universe.
When the universe becomes large, ECHDE converts to the
ordinary HDE \cite{Wei}.\\

For the first class (class I) of scale factor (15) and using
equation (16), the future event horizon $R_{h}$ (24) into (30) one
can get
\begin{equation}\label{31}
\rho_{\Lambda}=-\frac{T\delta^{2}(n+1)^{2}}{2n^{2}}+\frac{\alpha
T^{2}(n+1)^{2}}{36n^{4}}\ln(-\frac{6n^{2}}{T(n+1)^{2}})+\frac{\beta
T^{2}(n+1)^{4}}{36n^{4}}
\end{equation}
Substituting equation (31) in the differential equation (12) i.e.,
$\rho=\rho_{\Lambda}$, gives the following solution
\begin{equation}\label{32}
f(T)=c\sqrt{-T}-\frac{T^{2}(n+1)^{4}(3\beta+2\alpha)}{162n^{4}}-
\frac{\alpha(n+1)^{4}}{54n^{4}}T^{2}\ln(-\frac{6n^{2}}{T(n+1)^{2}})-\frac{\delta^{2}(n+1)^{2}}{n^{2}}
\end{equation}
where $c$ is the integration constant to be determined from the
necessary boundary condition. \textbf{In figure {\bf 7}}, we
understand that $f(T)\nrightarrow 0$ as $T\rightarrow 0$ for the
solution obtained from equation (32). We also observe that $f(T)$
first increases and then decreases as $T$ increases keeping in the
mind that $f(T)$ takes always negative value for all values of
negative $T$. Replacing equation (32) into (13) and using (31) we
obtain the EoS parameter of the ECHDE $f(T)$ gravity model in
logarithmic version as
$w_{\Lambda}=\frac{p_{\Lambda}}{\rho_{\Lambda}}$ graphically. In
figures {\bf 9} and {\bf 11}, we see that the EoS parameter can
justify the transition from phantom state $w_{\Lambda}<-1$, to the
quintessence regime, $w_{\Lambda}>-1$, i.e., it crosses the
phantom divide line $w_{\Lambda}=-1$ if we draw the graph of EoS
parameter with $T$ and $z$ using the equation (21) respectively.
So in this case, $f(T)$ gravity generates phantom crossing.\\

For the second class (class II) of scale factor (18) and using
equation (19), the future event horizon $R_{h}$ (27) into (30) one
can get
\begin{equation}\label{33}
\rho_{\Lambda}=-\frac{T\delta^{2}(n-1)^{2}}{2n^{2}}+\frac{\alpha
T^{2}(n-1)^{2}}{36n^{4}}\ln(-\frac{6n^{2}}{T(n-1)^{2}})+\frac{\beta
T^{2}(n-1)^{4}}{36n^{4}}
\end{equation}
Substituting equation (33) in the differential equation (12) i.e.,
$\rho=\rho_{\Lambda}$, gives the following solution
\begin{equation}\label{34}
f(T)=c\sqrt{-T}-\frac{T^{2}(n-1)^{4}(3\beta+2\alpha)}{162n^{4}}
-\frac{\alpha(n-1)^{4}}{54n^{4}}T^{2}\ln(-\frac{6n^{2}}{T(n-1)^{2}})-\frac{\delta^{2}(n-1)^{2}}{n^{2}}
\end{equation}
where $c$ is the integration constant to be determined from the
necessary boundary condition. \textbf{In figure {\bf 8}}, we
understand that $f(T)\nrightarrow0$ as $T\rightarrow0$ for the
solution obtained from equation (34). We also observe that $f(T)$
decreases from some positive value to negative value as $T$
increases from negative value to zero. It may be stated that the
solutions obtained in equations (32) and (34) are not so-realistic
models. Replacing equation (34) into (13) and using (33) we obtain
the EoS parameter of ECHDE $f(T)$ gravity model in logarithmic
version as $w_{\Lambda}=\frac{p_{\Lambda}}{\rho_{\Lambda}}$
graphically. In figures {\bf 10} and {\bf 12}, we see that the EoS
parameter can justify the transition from quintessence state
$w_{\Lambda}>-1$, to the phantom regime, $w_{\Lambda}<-1$, i.e.,
it crosses the phantom divide line $w_{\Lambda}=-1$ if we draw the
graph of EoS parameter with $T$ and $z$ using the equation (21) respectively
i.e., it crosses the line $w_{\Lambda}=-1$.\\

\begin{figure}

\includegraphics[height=2.0in]{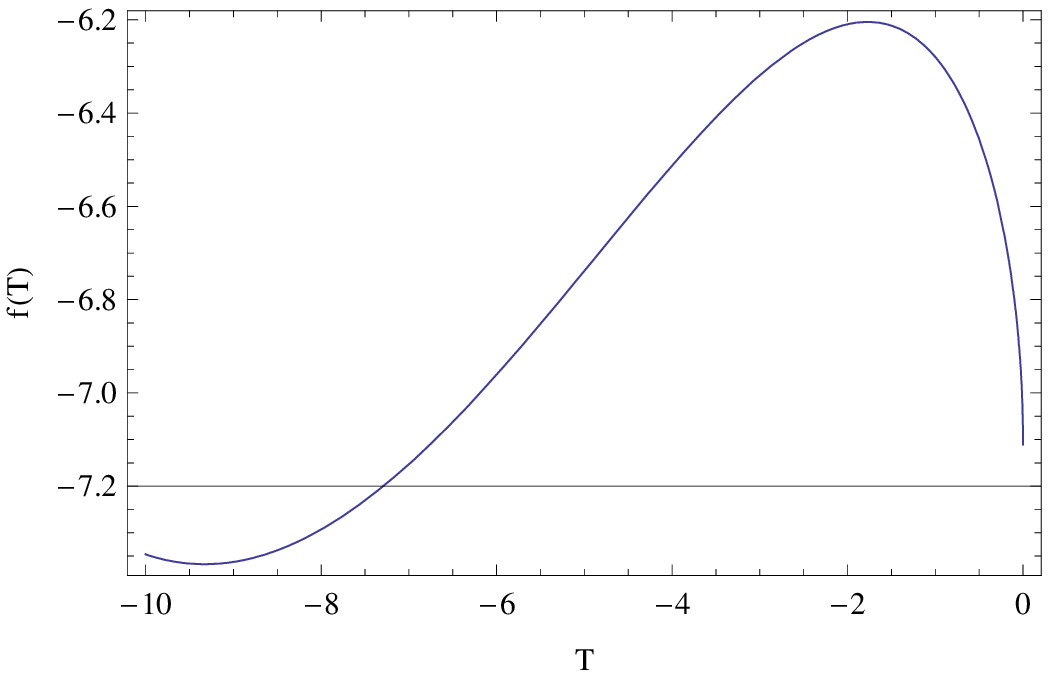}~~~~~
\includegraphics[height=2.0in]{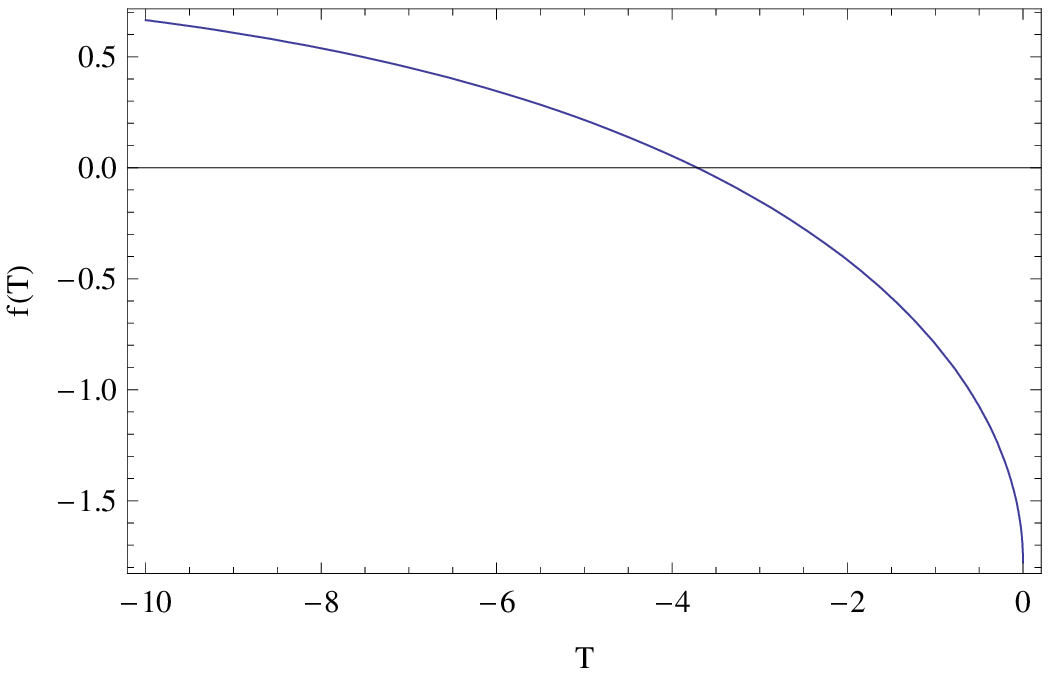}
\vspace{4 mm}
~~~~~~~~~~~~~~~~~~~~~~~Fig.7 ~~~~~~~~~~~~~~~~~~~~~~~~~~~~~~~~~~~~~~~~~~~~~~~~~~~~~~~~~~~~~~~~~~~~~~~Fig.8 \\
\vspace{4 mm}

\includegraphics[height=2.0in]{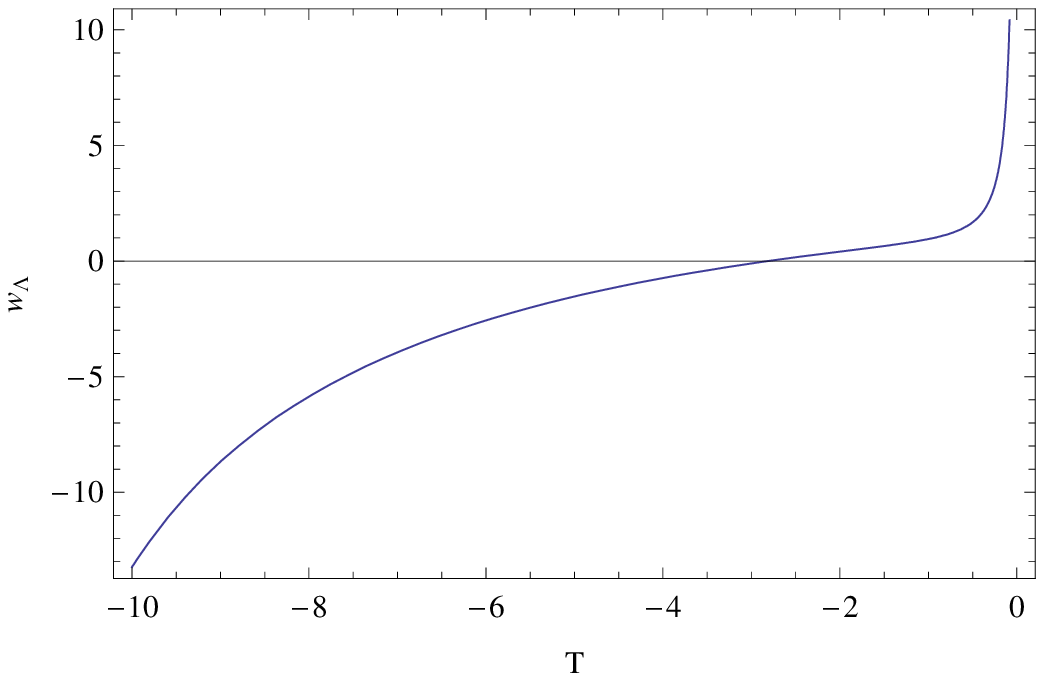}~~~~~
\includegraphics[height=2.0in]{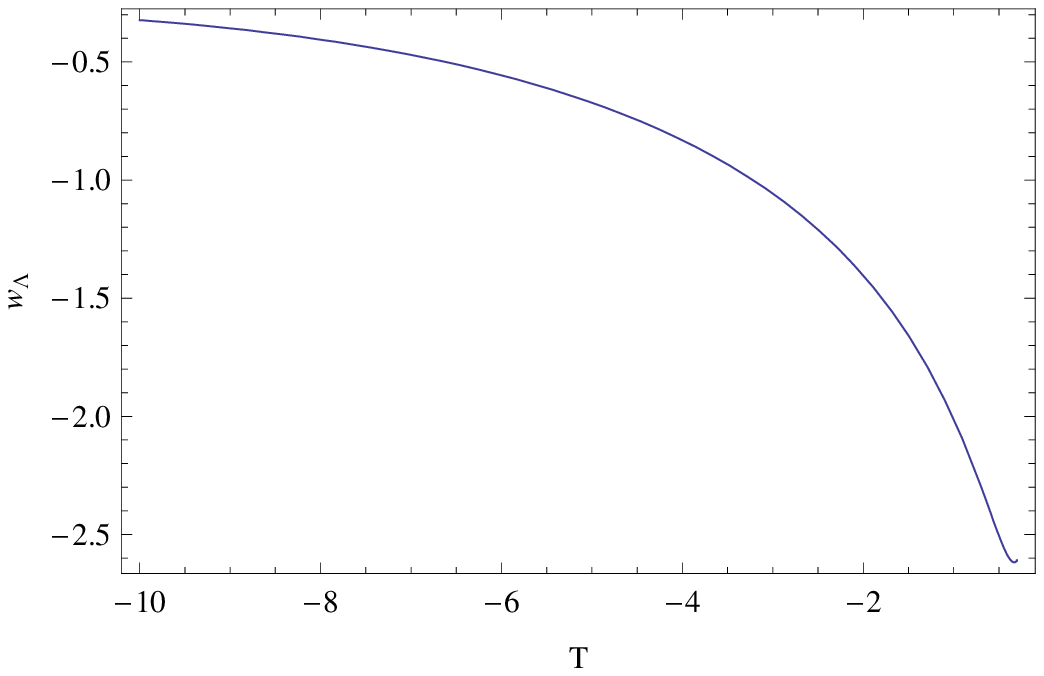}
\vspace{4 mm}
~~~~~~~~~~~~~~~~~~~~~~~Fig.9 ~~~~~~~~~~~~~~~~~~~~~~~~~~~~~~~~~~~~~~~~~~~~~~~~~~~~~~~~~~~~~~~~~~~~~~~Fig.10 \\
\vspace{4 mm}

\includegraphics[height=2.0in]{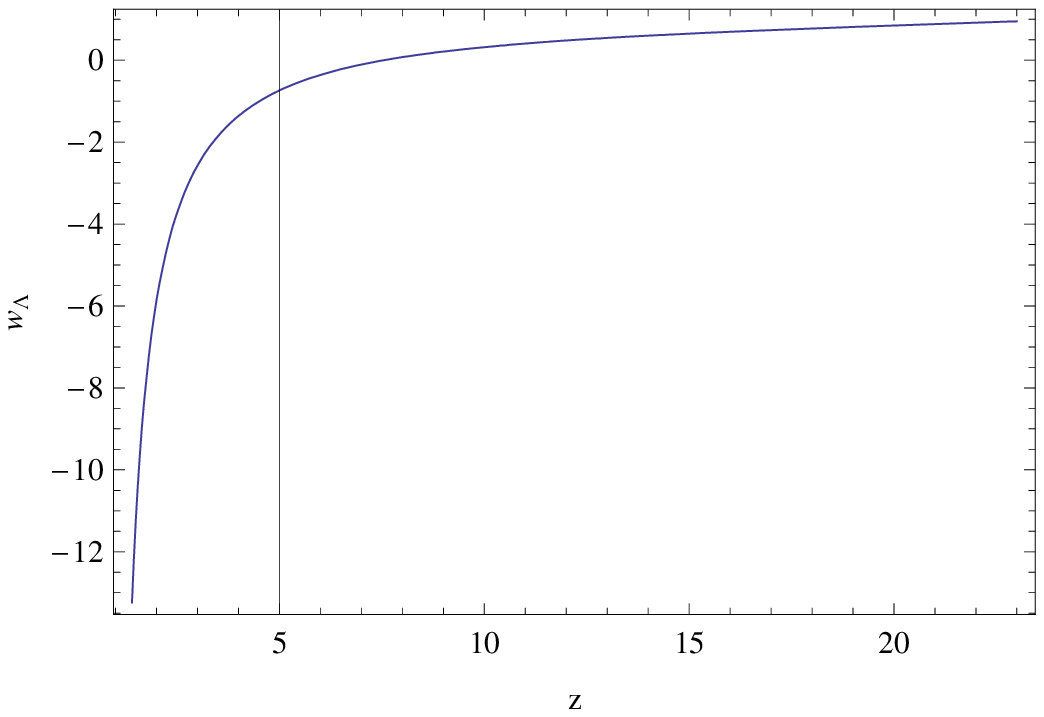}~~~~~~~~~~~~~~~~~~~~~~~~~~~~~~~~
\includegraphics[height=2.0in]{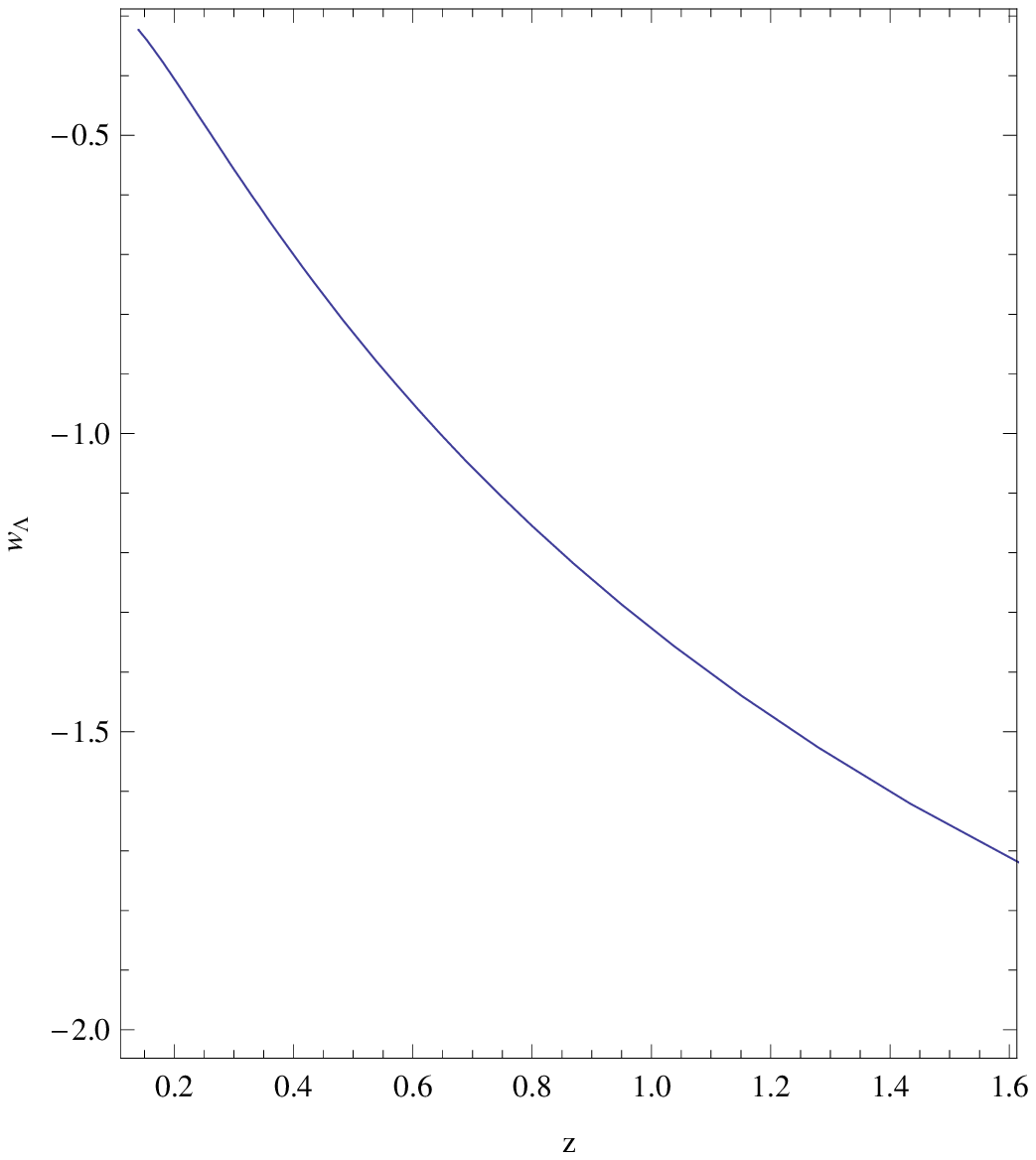}\\
\vspace{4 mm}
~~~~~~~~~~~~~~~~~~~~~~~Fig.11 ~~~~~~~~~~~~~~~~~~~~~~~~~~~~~~~~~~~~~~~~~~~~~~~~~~~~~~~~~~~~~~~~~~~~~~~Fig.12 \\
\vspace{4 mm}\\
\textbf{Figs.7 , 9 and 11} represent the plots of $f(T)$ and
$w_{\Lambda}$ for class I scale factor in ECHDE $f(T)$ gravity in
logarithmic correction model. \textbf{Figs.8 , 10 and 12}
represent the plots of $f(T)$ and $w_{\Lambda}$ for class II scale
factor in ECHDE $f(T)$ gravity in logarithmic correction model.
\end{figure}

\section{$f(T)$ reconstruction from ECNADE model in power-law and logarithmic corrections models}

\subsection{ECNADE in power-law correction}

\cite{Wei} gives the energy density of the ECNADE in power-law
correction with the help of quantum corrections to the relation
(1) in the setup of LQG given as

\begin{equation}\label{35}
\rho_{\Lambda}=3\delta^{2}\eta^{-2}-\lambda \eta^{-\epsilon}
\end{equation}
which are very similar to that of ECHDE in power-law correction
density (22) and $R_{h}$ is replaced with the conformal time
$\eta$ which is given by
\begin{equation}\label{36}
\eta=\int\frac{dt}{a}=\int\frac{da}{Ha^{2}}
\end{equation}
Here $\xi$ and $\zeta$ are dimensionless constants of order
unity.\\

For the first class (class I) of scale factor (15), the conformal
time $\eta$ by the help of equation (36) yields
\begin{equation}\label{37}
\eta=\int_{t}^{t_{s}}\frac{dt}{a}=-\frac{(t_{s}-t)^{1+n}}{a_{0}(1+n)}
=\sqrt{\frac{(6n^{2})^{n+1}}{a_{0}^{2}(-T)^{n+1}(1+n)^{2}}}
\end{equation}
Substituting equation (37) into (35) one can obtain

\begin{equation}\label{38}
\rho_{\Lambda}=\frac{3\delta^{2}a_{0}^{2}(n+1)^{2}(-T)^{n+1}}{(6n^{2})^{n+1}}-\lambda
\{\frac{a_{0}(n+1)}{6^{\frac{n+1}{2}}n^{n+1}}\}^{\epsilon}(-T)^{\frac{(n+1)\epsilon}{2}}
\end{equation}

Solving the differential equation (12) for the energy density (38)
reduces to i.e., $\rho=\rho_{\Lambda}$, gives the following
solution

\begin{equation}\label{39}
f(T)=c\sqrt{-T}-\frac{\delta^{2}a_{0}^{2}(n+1)^{2}(-T)^{n+1}}{6^{n}n^{2(n+1)}(2n+1)}+
\frac{2\lambda}{(n+1)\epsilon-1}\{\frac{a_{0}(n+1)}{6^{\frac{n+1}{2}}n^{n+1}}\}
^{\epsilon}(-T)^{\frac{(n+1)\epsilon}{2}}
\end{equation}

where $c$ is the integration constant to be determined from the
necessary boundary condition. \textbf{In figure {\bf 13}}, we
understand that $f(T)\nrightarrow 0$ as $T\rightarrow 0$ for the
solution obtained from equation (39). The function $f(T)$
decreases as $T$ increases to zero. Replacing equation (39) into
(13) and using (38) we obtain the EoS parameter of the ECNADE
$f(T)$ gravity model in power-law version as
$w_{\Lambda}=\frac{p_{\Lambda}}{\rho_{\Lambda}}$ graphically. In
figures {\bf 15} and {\bf 17}, we see that the EoS parameter can
justify the transition from phantom state $w_{\Lambda}<-1$, to the
quintessence regime, $w_{\Lambda}>-1$, i.e., it crosses the
phantom divide line $w_{\Lambda}=-1$ if we draw the graph of EoS
parameter with $T$ and $z$ using the equation (21) respectively
i.e., it crosses the line $w_{\Lambda}=-1$.\\

For the second class (class II) of scale factor (18), the
conformal time $\eta$ by the help of equation (36) yields
\begin{equation}\label{40}
\eta=\int_{0}^{t}\frac{dt}{a}=\frac{t^{1-n}}{a_{0}(1-n)}=\sqrt{\frac{6^{1-n}n^{2(1-n)}}{(-T)^{1-n}a_{0}^{2}(1-n)^{2}}}
\end{equation}
where $n<1$. Substituting the equation (40) into (35) one can
obtain
\begin{equation}\label{41}
\rho_{\Lambda}=\frac{3\delta^{2}a_{0}^{2}(1-n)^{2}(-T)^{1-n}}{(6n^{2})^{1-n}}-\lambda
\{\frac{a_{0}(1-n)}{6^{\frac{1-n}{2}}n^{1-n}}\}^{\epsilon}(-T)^{\frac{(1-n)\epsilon}{2}}
\end{equation}
Solving the differential equation (12) for the energy density (41)
reduces to i.e., $\rho=\rho_{\Lambda}$, gives the following
solution
\begin{equation}\label{42}
f(T)=c\sqrt{-T}-\frac{6^{n}\delta^{2}a_{0}^{2}(1-n)^{2}(-T)^{1-n}}{n^{2(1-n)}(1-2n)}
+\frac{2\lambda}{(1-n)\epsilon-1}\{\frac{a_{0}(1-n)}{6^{\frac{1-n}{2}}n^{1-n}}\}
^{\epsilon}(-T)^{\frac{(1-n)\epsilon}{2}}
\end{equation}

where $c$ is the integration constant to be determined from the
necessary boundary condition. \textbf{In figure {\bf 14}}, we
understand that $f(T)\nrightarrow 0$ as $T\rightarrow 0$ for the
solution obtained from equation (42). The function $f(T)$
decreases but keeps negative value as $T$ increases to zero.It may
be stated that the solutions obtained in equation (39) and (42)
both are not realistic models. Replacing equation (42) into (13)
and using (41) we obtain the EoS parameter of the ECNADE $f(T)$
gravity model in power-law version as
$w_{\Lambda}=\frac{p_{\Lambda}}{\rho_{\Lambda}}$ graphically. In
figures {\bf 16} and {\bf 18}, we see that the EoS parameter
entirely lies in the phantom region, $w_{\Lambda}<-1$ if we draw
the graph of EoS parameter with $T$ and $z$ using the equation
(21) respectively. So in this case, $f(T)$ gravity generates
phantom crossing i.e., it does not cross the line $w_{\Lambda}=-1$.\\

\begin{figure}

\includegraphics[height=2.0in]{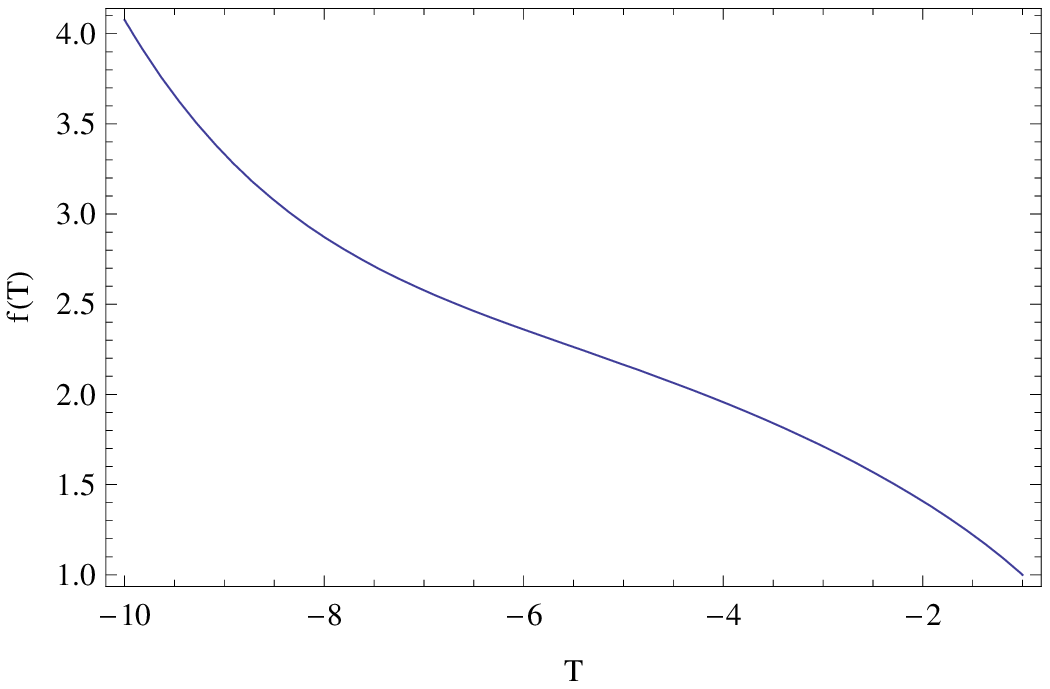}~~~~~
\includegraphics[height=2.0in]{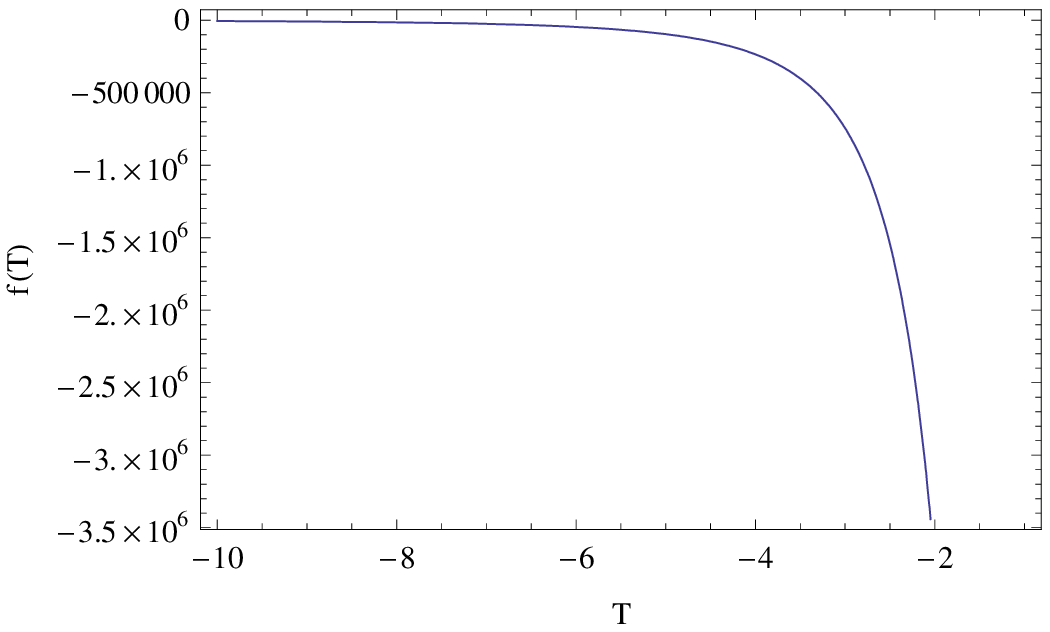}
\vspace{4 mm}
~~~~~~~~~~~~~~~~~~~~~~~Fig.13 ~~~~~~~~~~~~~~~~~~~~~~~~~~~~~~~~~~~~~~~~~~~~~~~~~~~~~~~~~~~~~~~~~~~~~~~Fig.14 \\
\vspace{4 mm}

\includegraphics[height=2.0in]{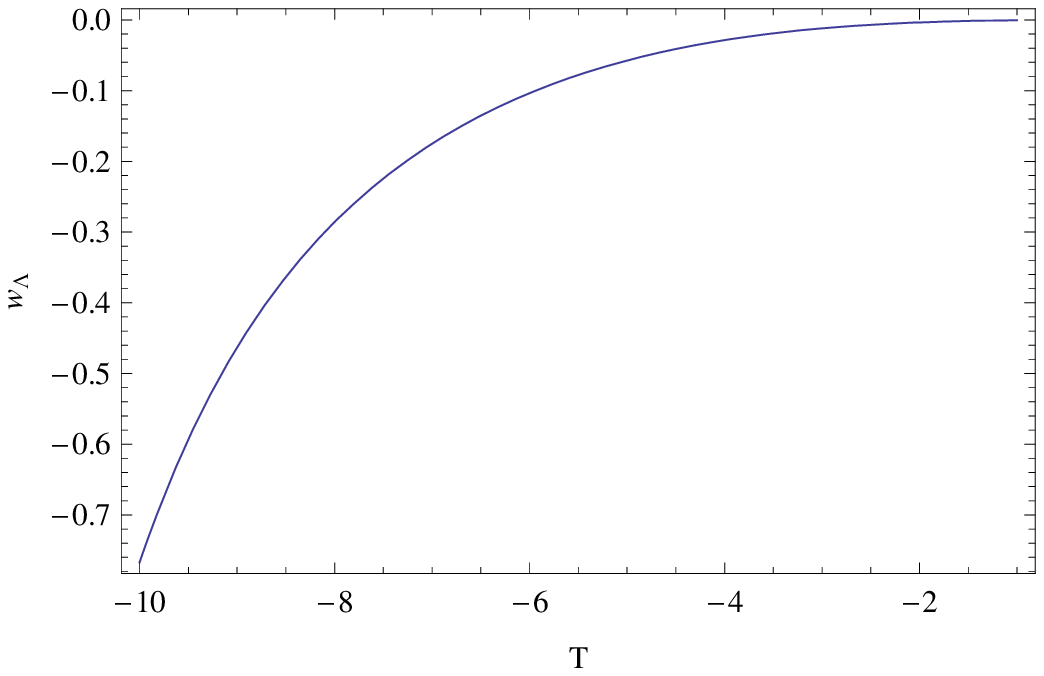}~~~~~
\includegraphics[height=2.0in]{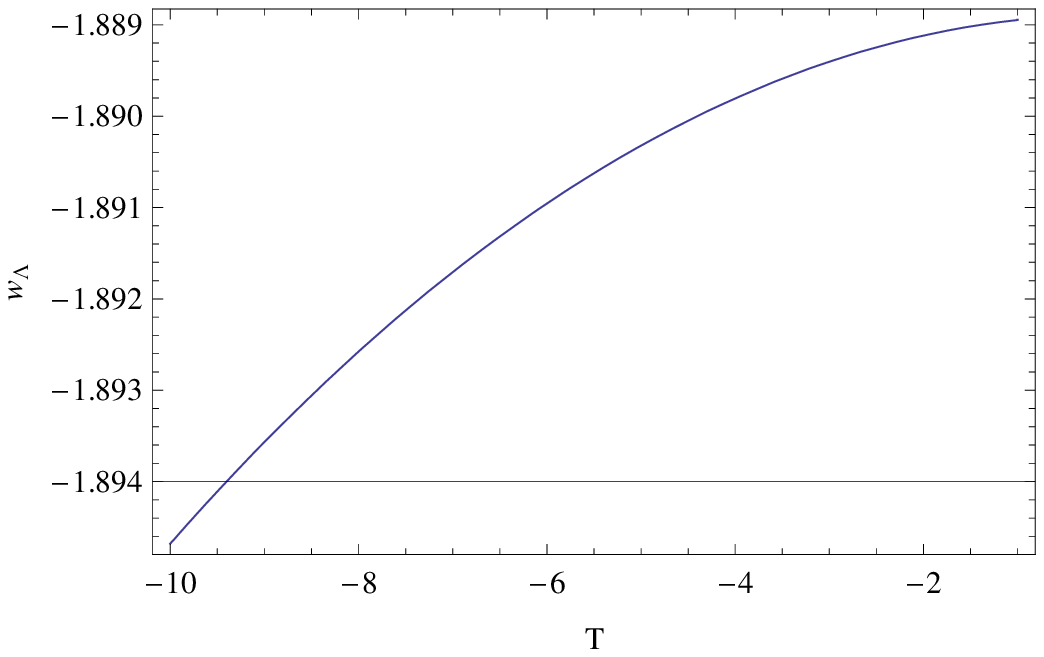}
\vspace{4 mm}
~~~~~~~~~~~~~~~~~~~~~~~Fig.15 ~~~~~~~~~~~~~~~~~~~~~~~~~~~~~~~~~~~~~~~~~~~~~~~~~~~~~~~~~~~~~~~~~~~~~~~Fig.16 \\
\vspace{4 mm}

\includegraphics[height=2.0in]{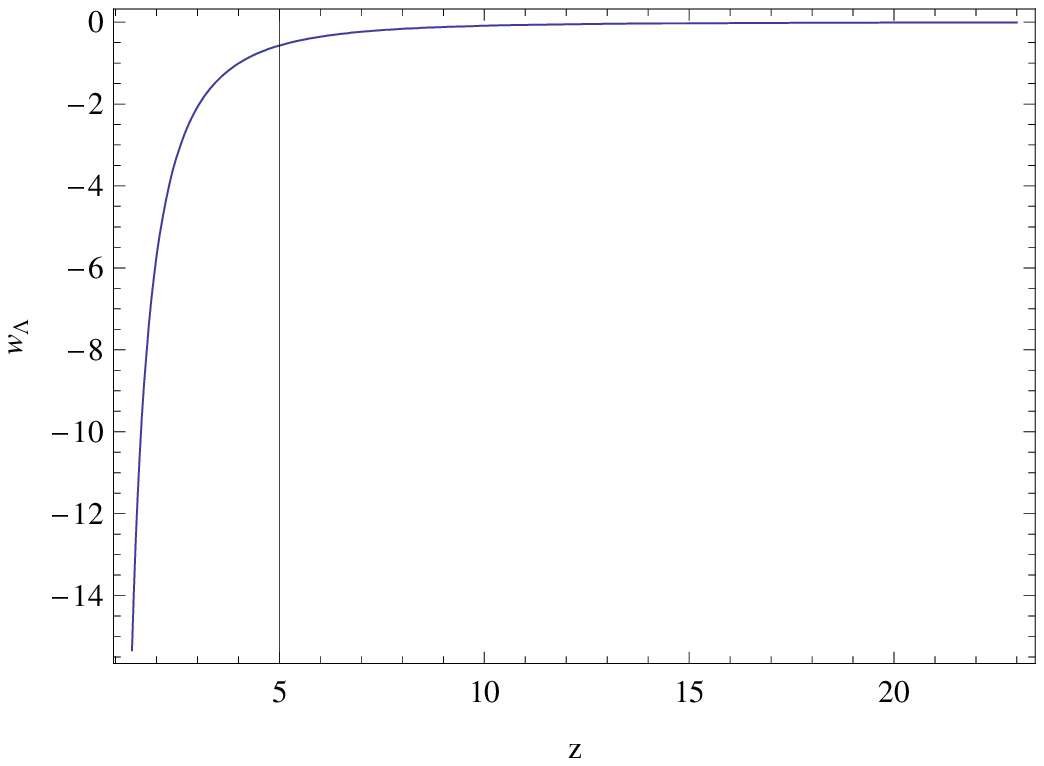}~~~~~
\includegraphics[height=2.0in]{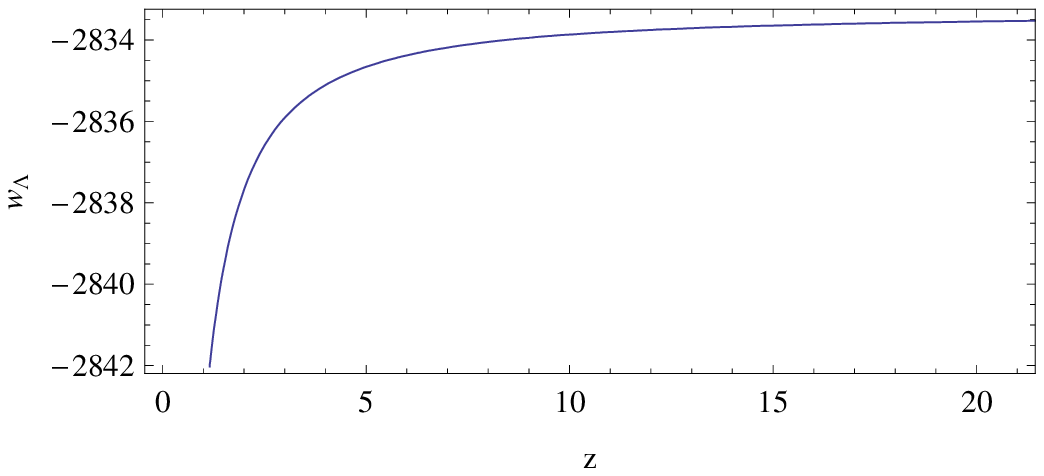}
\vspace{4 mm}
~~~~~~~~~~~~~~~~~~~~~~~Fig.17 ~~~~~~~~~~~~~~~~~~~~~~~~~~~~~~~~~~~~~~~~~~~~~~~~~~~~~~~~~~~~~~~~~~~~~~~Fig.18 \\
\vspace{4 mm}\\
 \textbf{Figs.13, 15 and 17} represent the plots of $f(T)$ and
$w_{\Lambda}$ for class I scale factor in ECNADE $f(T)$ gravity
power-law correction model. \textbf{Figs.14, 16 and 18} represent
the plots of $f(T)$ and $w_{\Lambda}$ for class II scale factor in
ECNADE $f(T)$ gravity in power-law correction model.
\end{figure}

\subsection{ECNADE in logarithmic correction}

\cite{Wei} gives the energy density of the ECNADE with the help of
quantum corrections to the entropy-area relation (2) in the setup
of LQG given as
\begin{equation}\label{43}
\rho_{\Lambda}=\frac{3\alpha^{2}}{\eta^{2}}+\frac{\xi}{\eta^{4}}\ln(\eta^{2})+\frac{\zeta}{\eta^{4}}
\end{equation}
which are very similar to that of ECHDE density in logarithmic
version (30) and $R_{h}$ is replaced with the conformal time
$\eta$.\\

For the first class (class I) of scale factor (15), using the
conformal time $\eta$ (40), equation (43) gives
\begin{equation}\label{44}
\rho_{\Lambda}=\frac{3\alpha^{2}a_{0}^{2}(1+n)^{2}(-T)^{n+1}}{(6n^{2})^{n+1}}+\frac{\xi
a_{0}^{4}(1+n)^{4}(-T)^{2n+2}}{(6n^{2})^{2n+2}}\ln(\frac{(6n^{2})^{n+1}}{a_{0}^{2}(1+n)^{2}(-T)^{n+1}})+\frac{\zeta
a_{0}^{4}(1+n)^{4}(-T)^{2n+2}}{(6n^{2})^{2n+2}}
\end{equation}
Solving the differential equation (12) for the energy density (44)
reduces to i.e., $\rho=\rho_{\Lambda}$, gives the following
solution

\begin{eqnarray*}
f(T)=c\sqrt{-T}-\frac{3\alpha^{2}(-T)^{n+1}a_{0}^{2}(1+n)^{2}}{(6n^{2})^{n+1}(n+\frac{1}{2})}-\frac{\xi
a_{0}^{4}(1+n)^{4}(-T)^{2n+2}}{(6n^{2})^{2n+2}(2n+\frac{3}{2})}
\ln(\frac{(6n^{2})^{n+1}}{a_{0}^{2}(n+1)^{2}(-T)^{n+1}})
\end{eqnarray*}
\begin{equation}\label{45}
+\frac{(-T)^{2n+4}}{n(2n+\frac{3}{2})(2n+\frac{7}{2})})
-\frac{\zeta
T^{2n+2}a_{0}^{4}(1+n)^{4}}{(6n^{2})^{2n+2}(2n+\frac{3}{2})}
\end{equation}
where $c$ is the integration constant to be determined from the
necessary boundary condition. \textbf{In figure {\bf 19}}, we
understand that $f(T)\rightarrow 0$ as $T\rightarrow 0$ for the
solution obtained from equation (45). The function $f(T)$
increases but keeps negative value as $T$ increases to zero.
Replacing equation (45) into (13) and using (44) we obtain the EoS
parameter of the ECNADE $f(T)$ gravity model in logarithmic
version as $w_{\Lambda}=\frac{p_{\Lambda}}{\rho_{\Lambda}}$
graphically. In figures {\bf 21} and {\bf 23}, we see that the EoS
parameter can justify the transition from quintessence state
$w_{\Lambda}>-1$, to the phantom regime, $w_{\Lambda}<-1$, i.e.,
it crosses the phantom divide line $w_{\Lambda}=-1$ if we draw the
graph of EoS parameter with $T$ and $z$ using the equation (21) respectively
i.e., it crosses the line $w_{\Lambda}=-1$.\\

For the second class (class II) of scale factor (18), the
conformal time $\eta$ (40) equation (43) gives
\begin{equation}\label{46}
\rho_{\Lambda}=\frac{3\alpha^{2}a_{0}^{2}(1-n)^{2}(-T)^{1-n}}{6^{1-n}n^{2(1-n)}}+\frac{\xi
a_{0}^{4}(1-n)^{4}T^{2-2n}}{6^{2(1-n)}n^{4(1-n)}}\ln(\frac{6^{1-n}n^{2(1-n)}}{a_{0}^{2}(1-n)^{2}(-T)^{1-n}})+\frac{\zeta
a_{0}^{4}(1-n)^{4}T^{2-2n}}{6^{2(1-n)}n^{4(1-n)}}
\end{equation}
Solving the differential equation (12) for the energy density (46)
reduces to
\begin{eqnarray*}
f(T)=c\sqrt{-T}-\frac{3\alpha^{2}a_{0}^{2}(1-n)^{2}(-T)^{1-n}}{6^{1-n}n^{2(1-n)}(\frac{1}{2}-n)}-\frac{\xi
a_{0}^{4}(1-n)^{4}T^{2-2n}}{6^{2(1-n)}n^{4(1-n)}(\frac{3}{2}-2n)^{2}}((\frac{3}{2}-2n)
\ln(\frac{6^{1-n}n^{2(1-n)}}{a_{0}^{2}(1-n)^{2}(-T)^{1-n}})+(1-n))
\end{eqnarray*}
\begin{equation}\label{47}
-\frac{\zeta
a_{0}^{4}(1-n)^{4}T^{2-2n}}{6^{2(1-n)}n^{4(1-n)}(\frac{3}{2}-2n)}
\end{equation}
where $c$ is the integration constant to be determined from the
necessary boundary condition. \textbf{In figure {\bf 20}}, we
understand that $f(T)\rightarrow0$ as $T\rightarrow0$ for the
solution obtained from equation (47). The function $f(T)$
decreases from some positive value to some negative value as $T$
increases upto certain negative value and after that $f(T)$
increases keeping in negative sign. It may be stated that the
solutions obtained in equation (45) and (47) are both realistic
model. Replacing equation (47) into (13) and using (46) we obtain
the EoS parameter of the ECNADE $f(T)$ gravity model in
logarithmic version as
$w_{\Lambda}=\frac{p_{\Lambda}}{\rho_{\Lambda}}$ graphically. In
figures {\bf 22} and {\bf 24}, we see that the EoS parameter can
justify the transition from quintessence state $w_{\Lambda}>-1$,
to the phantom regime, $w_{\Lambda}<-1$, i.e., it crosses the
phantom divide line $w_{\Lambda}=-1$ if we draw the graph of EoS
parameter with $T$ and $z$ using the equation (21) respectively
i.e., it crosses the line $w_{\Lambda}=-1$.\\

\begin{figure}

\includegraphics[height=2.0in]{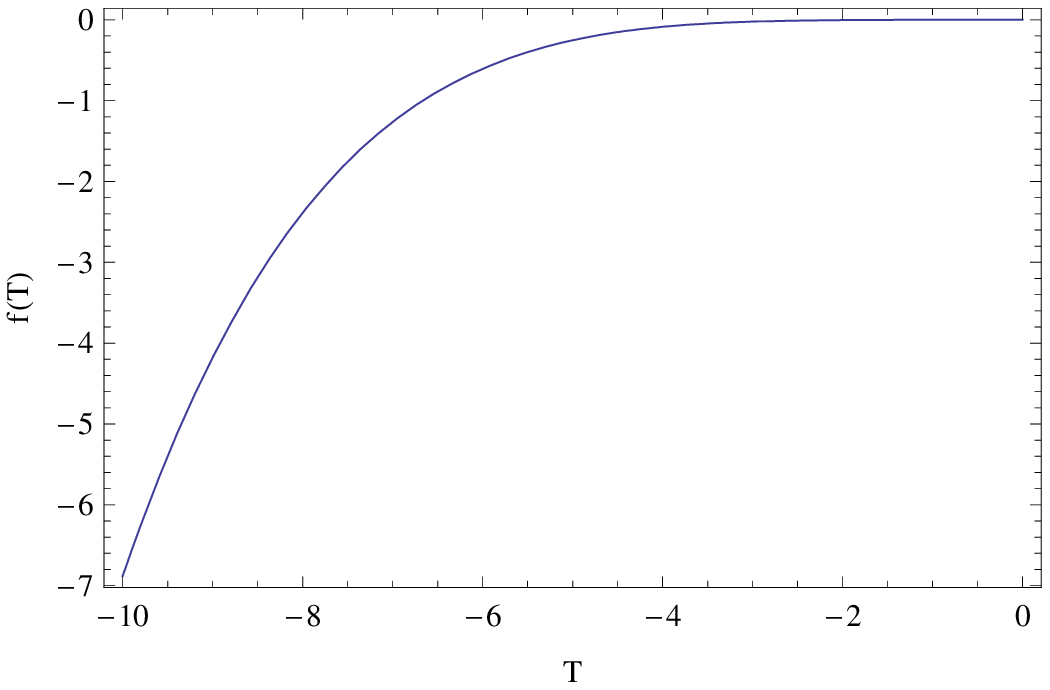}~~~~~
\includegraphics[height=2.0in]{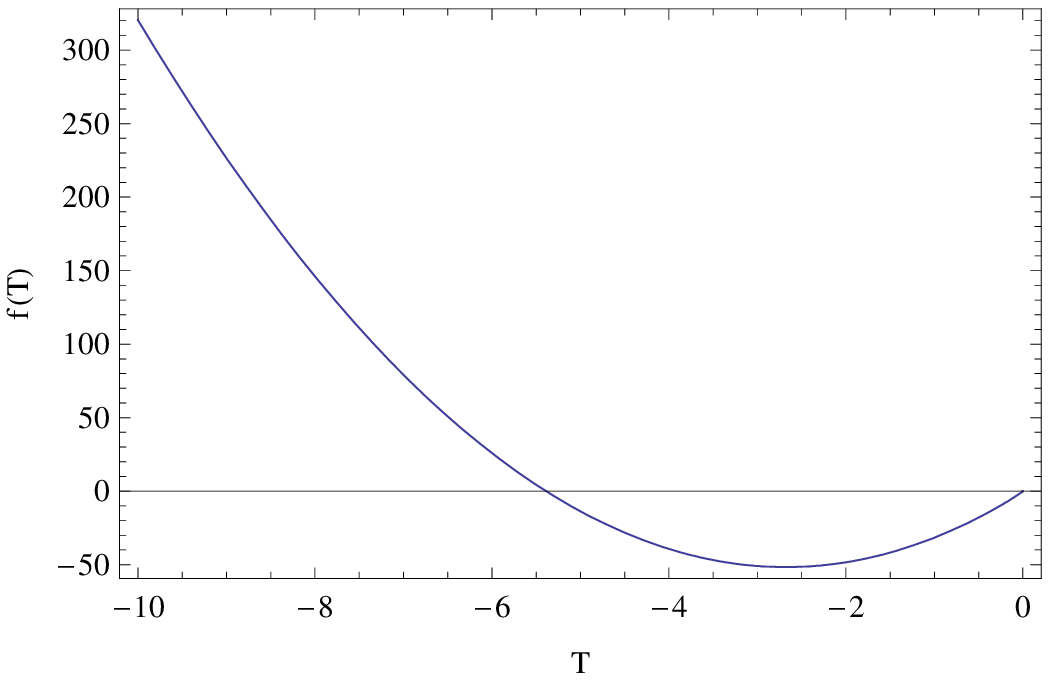}
\vspace{4 mm}
~~~~~~~~~~~~~~~~~~~~~~~Fig.19 ~~~~~~~~~~~~~~~~~~~~~~~~~~~~~~~~~~~~~~~~~~~~~~~~~~~~~~~~~~~~~~~~~~~~~~~Fig.20 \\
\vspace{4 mm}

\includegraphics[height=2.0in]{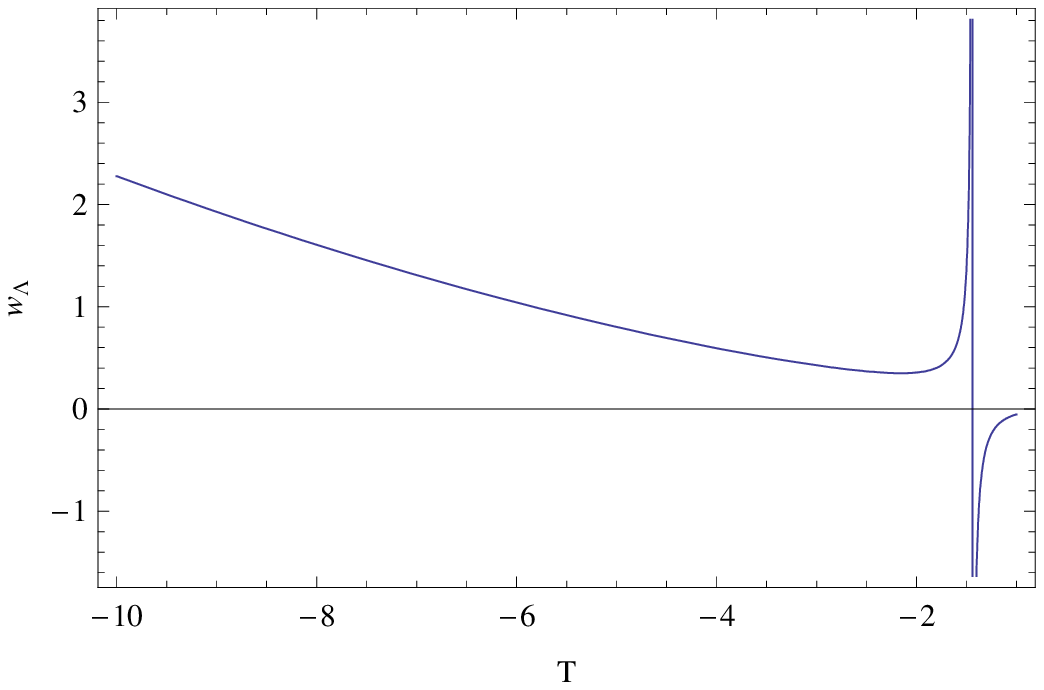}~~~~~
\includegraphics[height=2.0in]{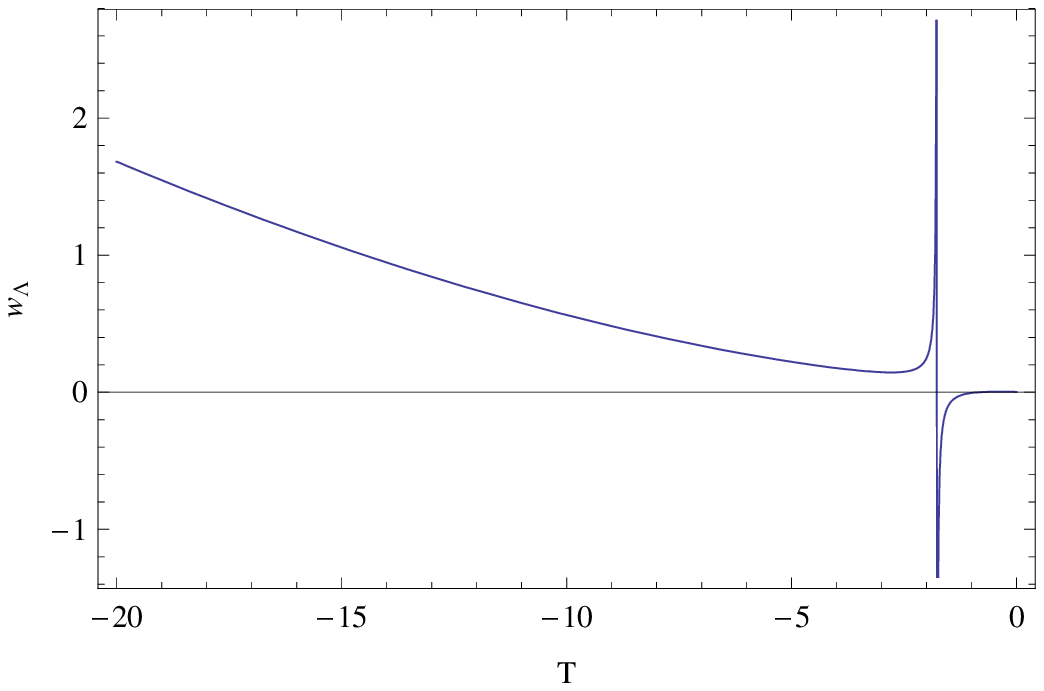}
\vspace{4 mm}
~~~~~~~~~~~~~~~~~~~~~~~Fig.21 ~~~~~~~~~~~~~~~~~~~~~~~~~~~~~~~~~~~~~~~~~~~~~~~~~~~~~~~~~~~~~~~~~~~~~~~Fig.22 \\
\vspace{4 mm}

\includegraphics[height=2.0in]{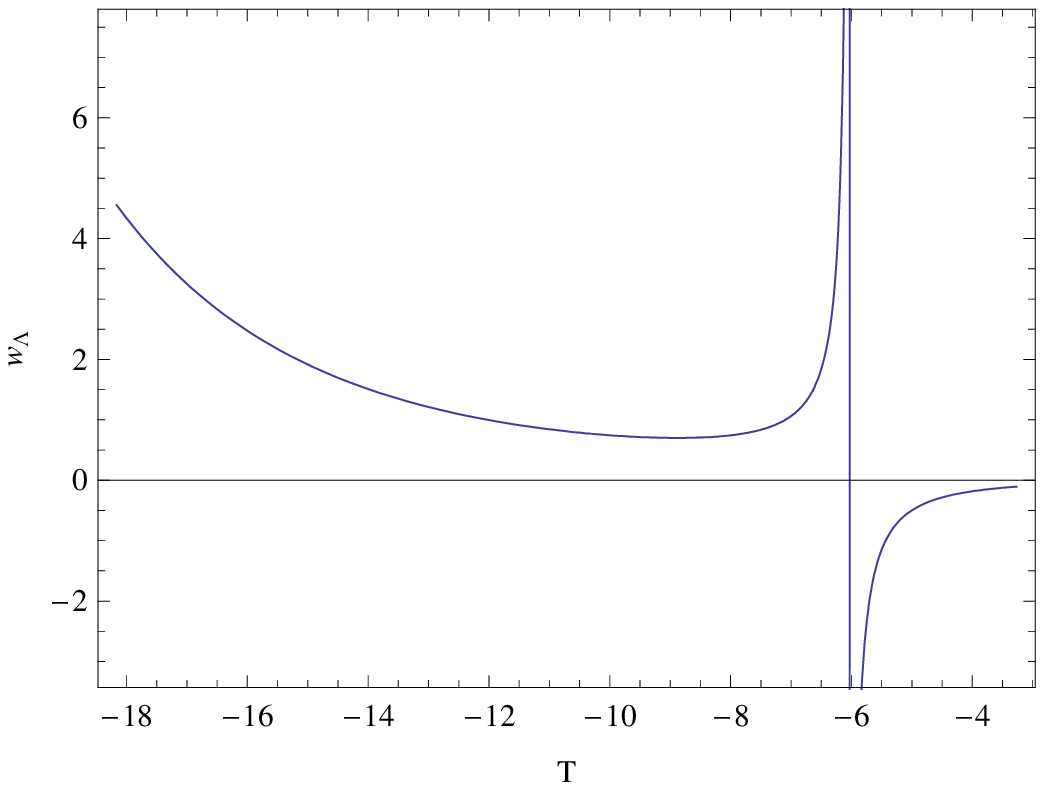}~~~~~
\includegraphics[height=2.0in]{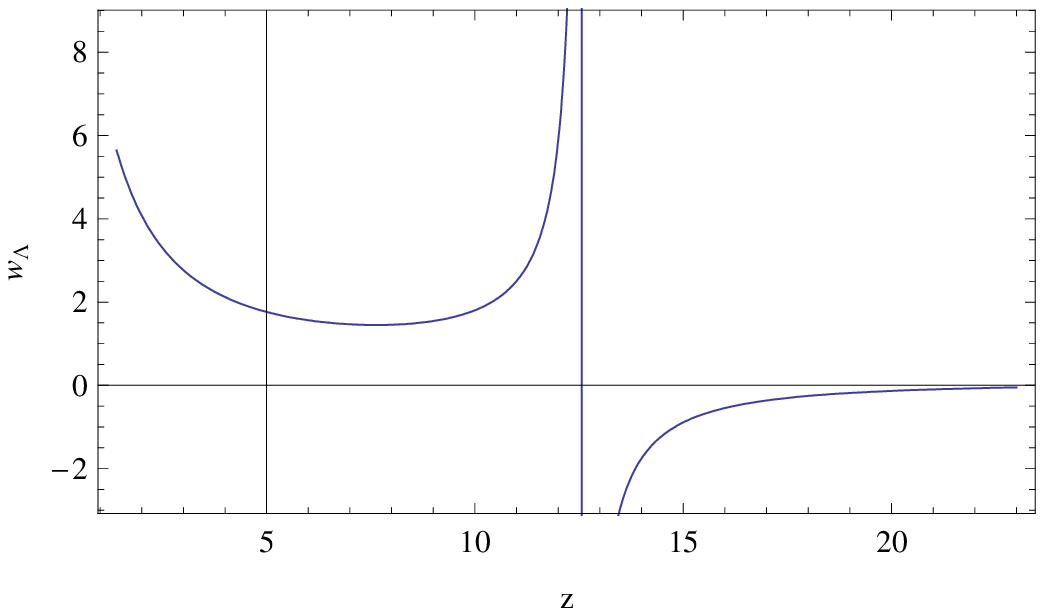}
\vspace{4 mm}
~~~~~~~~~~~~~~~~~~~~~~~Fig.23 ~~~~~~~~~~~~~~~~~~~~~~~~~~~~~~~~~~~~~~~~~~~~~~~~~~~~~~~~~~~~~~~~~~~~~~~Fig.24 \\
\vspace{4 mm}\\
 \textbf{Figs.19, 21 and 23} represent the plots of $f(T)$ and
$w_{\Lambda}$ for class I scale factor in ECNADE $f(T)$ gravity in
logarithmic correction model. \textbf{Figs.20, 22 and 24}
represent the plots of $f(T)$ and $w_{\Lambda}$ for class II scale
factor in ECNADE $f(T)$ gravity in logarithmic correction model.
\end{figure}

\section{Analysis and comparison of the reconstructed models}

We now analyze an important quantity to verify the stability of
ECHDE $f(T)$ in power-law and logarithmic corrections model and
ECNADE $f(T)$ in power-law and logarithmic corrections model,
named as the squared speed of sound $v_{s}^{2}$:
\begin{equation}\label{48}
v_{s}^{2}=\frac{dp}{d\rho}=\frac{\frac{dp}{dT}}{\frac{d\rho}{dT}}
\end{equation}
The sign of $v_{s}^{2}$ is very important for checking the
stability of a background evolution of the universe. In general
relativity a negative sign implies a classical instability of a
given perturbation \cite{Kim,Myung}. Myung \cite{Myung} has
observed the always negative sign of $v_{s}^{2}$ for HDE for the
future event horizon as IR cutoff, while for Chaplygin gas and
tachyon, there is non-negativity. Kim et al \cite{Kim} found
always negative squared speed of sound for agegraphic DE leading
to the instability of the perfect fluid for the model. Also,
\cite{Ebrahimi and Sheykhi} found the ghost QCD  DE model as
unstable model. Recently, Sharif and Jawad \cite{Sharif and Jawad}
have
shown negative $v_{s}^{2}$ for the interacting new HDE.\\

\subsection{Investigation of stability of ECHDE in power-law and logarithmic
corrections:}

For ECHDE $f(T)$ model in power-law version there are two cases.
For the first class (class I scale factor) we see from
\textbf{figure {\bf 25} that $v_{s}^{2}>0$ for $T \preceq -2$ and
$v_{s}^{2}<0$ for $T \succeq -2$ and from {\bf 27} that
$v_{s}^{2}<0$ for $z \preceq 0.1$ and $v_{s}^{2}>0$ for $z \succeq
0.1$ and for the second class (class II scale factor) we see from
figures {\bf 26} and {\bf 28}} that $v_{s}^{2}<0$ for the present
and future and future epoch. So we can conclude that ECHDE $f(T)$
model in power-law version implies \textbf{a classical stability
for $T \preceq -2$, $z \succeq 0.1$ and classically instability
for $T \succeq -2$, $z \preceq 0.1$} for the first class and a
classically instability of second class of a given perturbation in
general relativity.\\

For ECHDE $f(T)$ model in logarithmic version there are two cases.
For the first class (class I scale factor) we see from
\textbf{figure {\bf 29} and {\bf 31}} that $v_{s}^{2}<0$ for the
present and future epoch and for the second class (class II scale
factor) we see from \textbf{figure {\bf 30} and {\bf 32}} that
$v_{s}^{2}<0$ also for the present and future and future epoch. So
we can conclude that ECHDE $f(T)$ model in logarithmic version
implies a classical instability of a given perturbation in general
relativity for the
first and second classes both.\\

\begin{figure}

\includegraphics[height=2.0in]{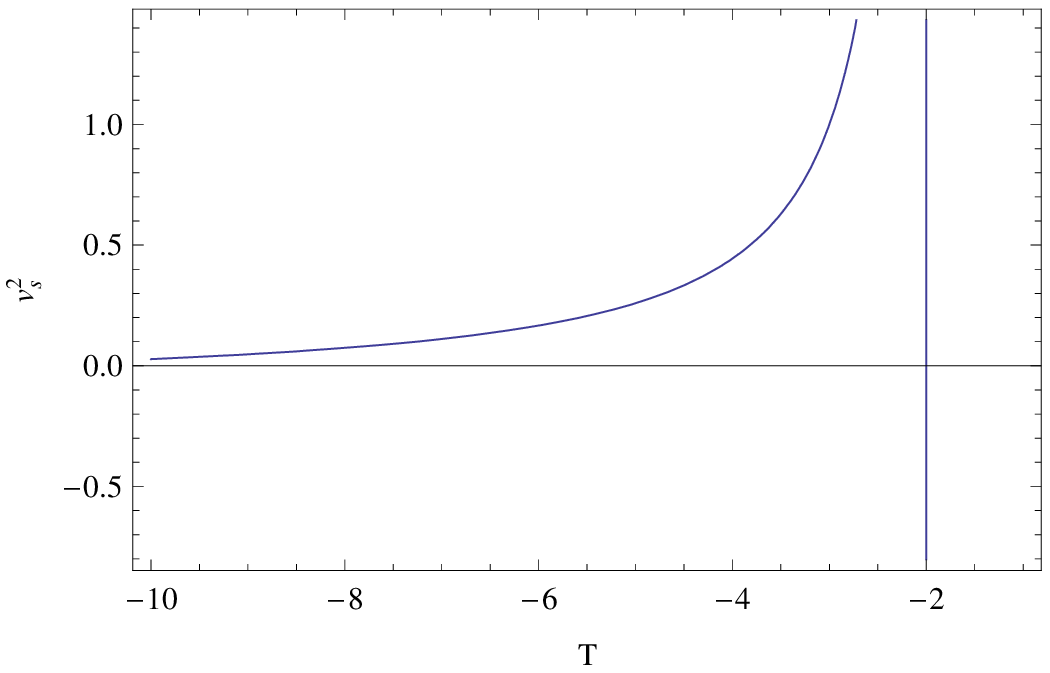}~~~~~
\includegraphics[height=2.0in]{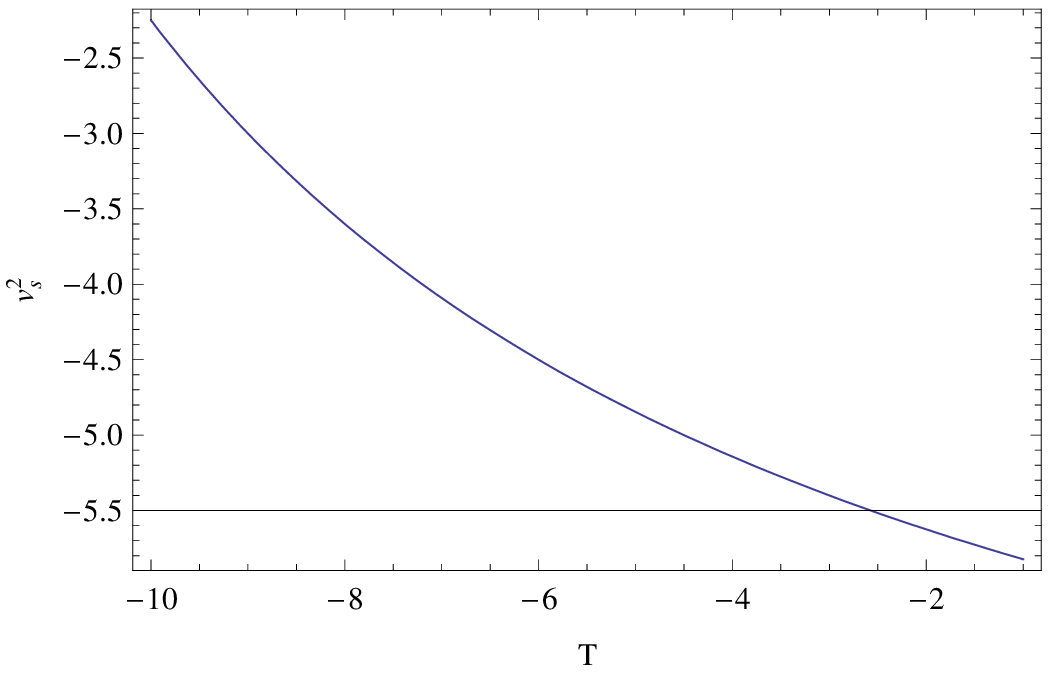}
\vspace{2 mm}
~~~~~~~~~~~~~~~~~~~~~~~Fig.25 ~~~~~~~~~~~~~~~~~~~~~~~~~~~~~~~~~~~~~~~~~~~~~~~~~~~~~~~~~~~~~~~~~~~~~~~Fig.26 \\
\vspace{1 mm}

\includegraphics[height=2.0in]{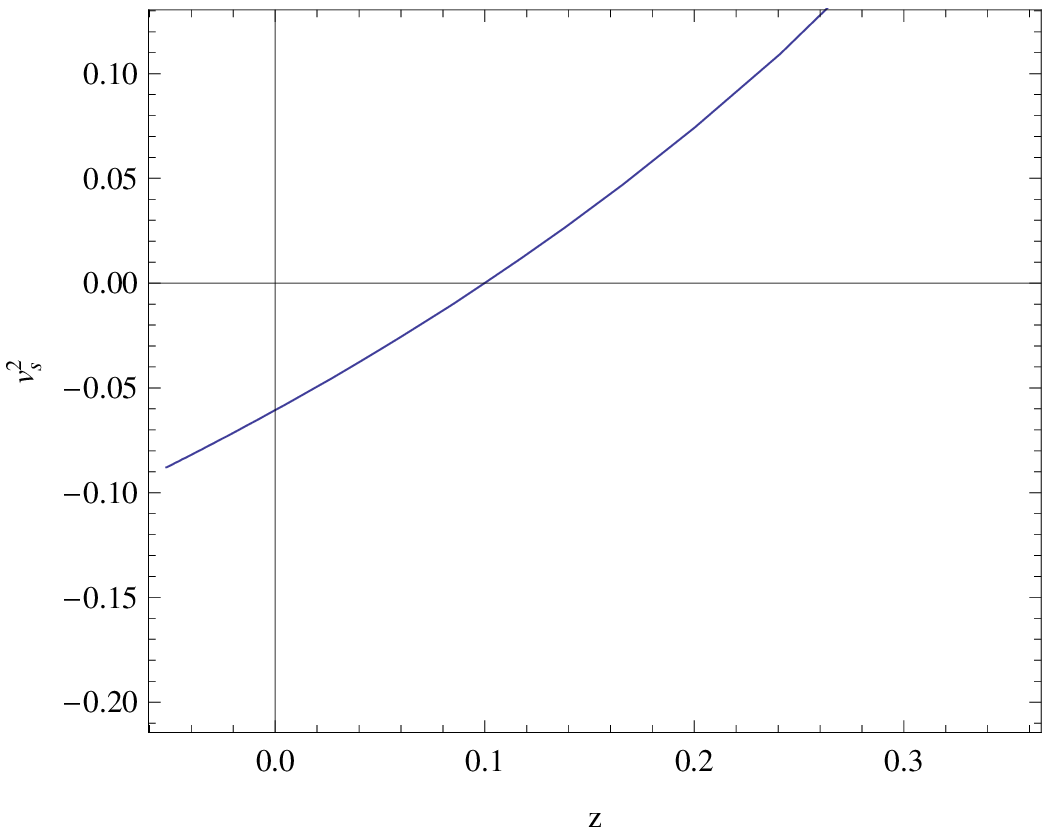}~~~~~~~~~~~~~~~~~~~~~~~~~~
\includegraphics[height=2.0in]{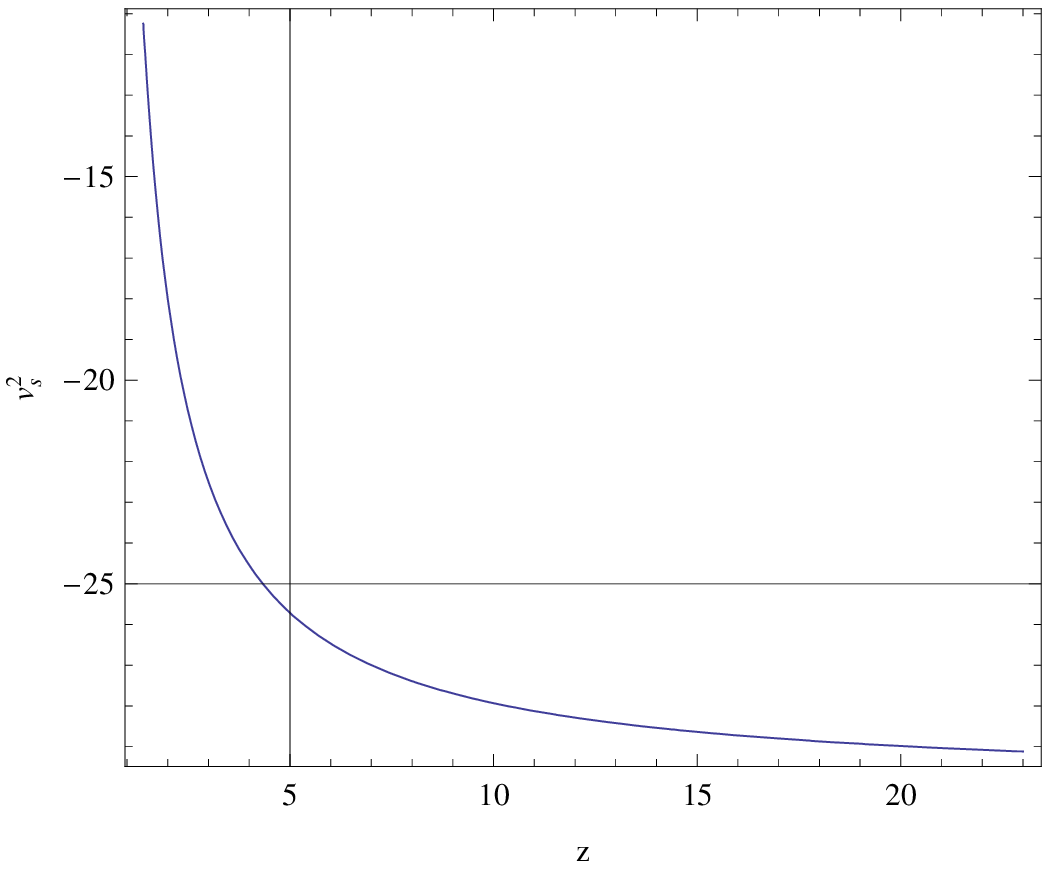}\\
\vspace{2 mm}
~~~~~~~~~~~~~~~~~~~~~~~Fig.27 ~~~~~~~~~~~~~~~~~~~~~~~~~~~~~~~~~~~~~~~~~~~~~~~~~~~~~~~~~~~~~~~~~~~~~~~Fig.28 \\
\vspace{1 mm}

\includegraphics[height=2.0in]{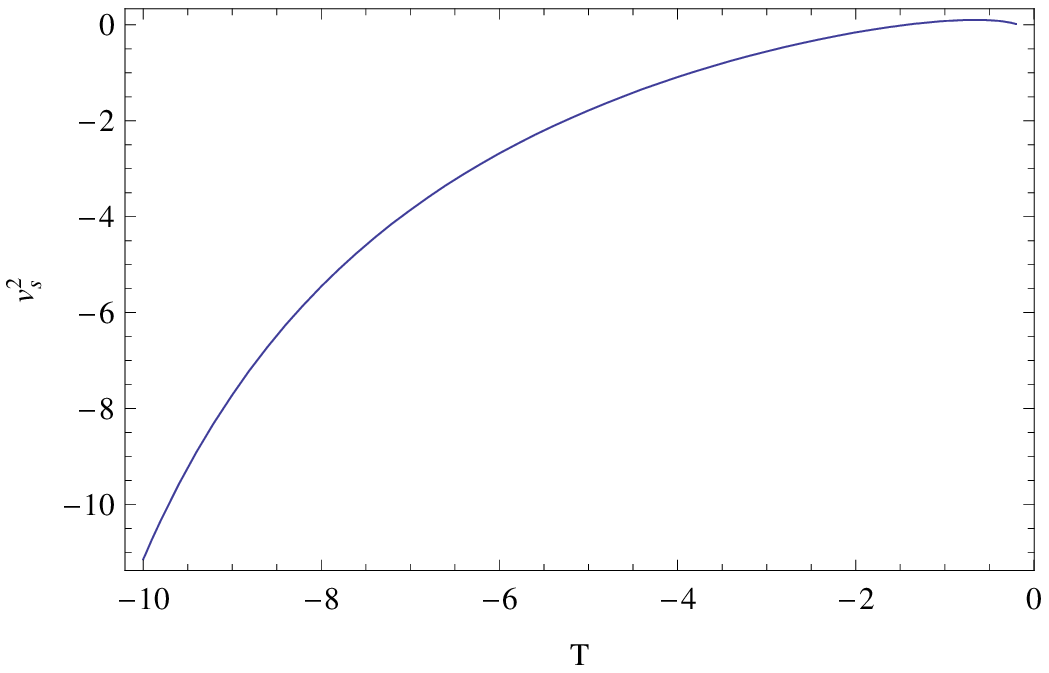}~~~~~
\includegraphics[height=2.0in]{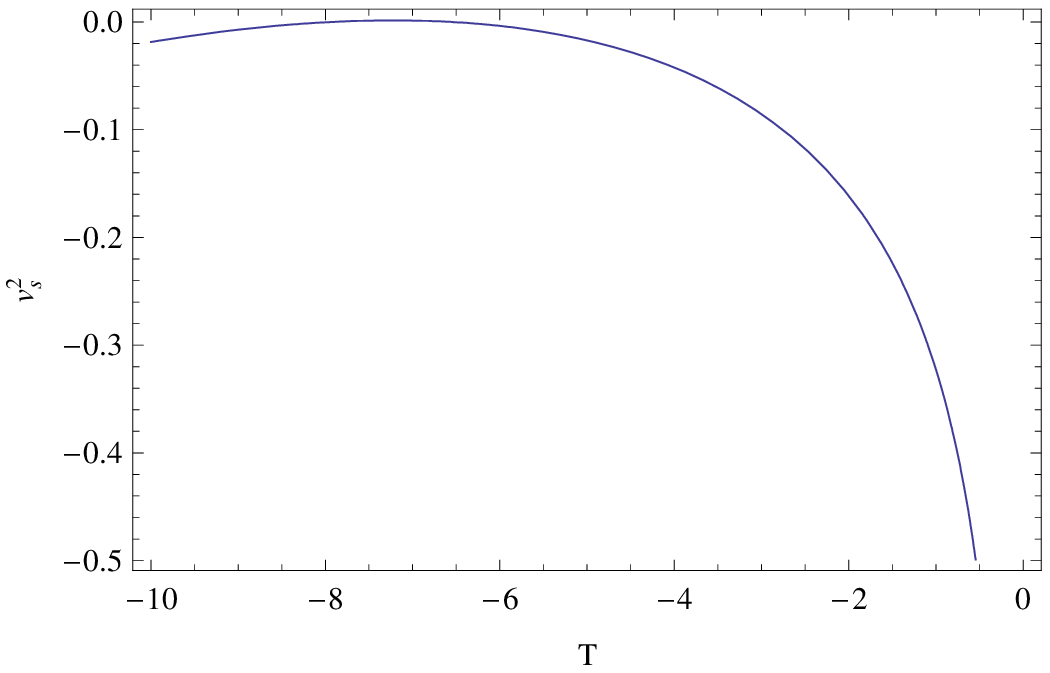}
\vspace{2 mm}
~~~~~~~~~~~~~~~~~~~~~~~Fig.29 ~~~~~~~~~~~~~~~~~~~~~~~~~~~~~~~~~~~~~~~~~~~~~~~~~~~~~~~~~~~~~~~~~~~~~~~Fig.30 \\
\vspace{1 mm}

\includegraphics[height=2.0in]{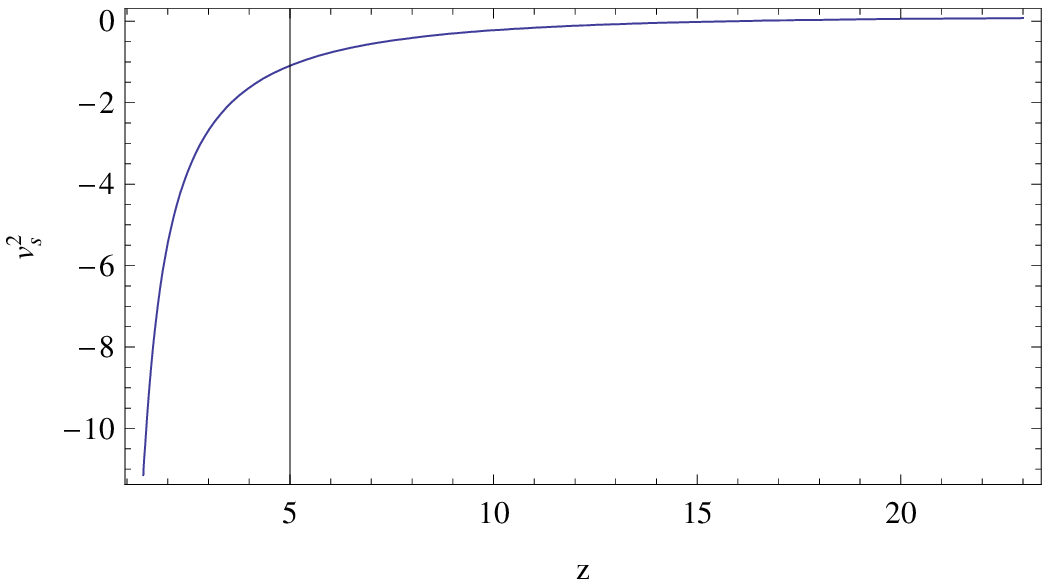}~~~~~
\includegraphics[height=2.0in]{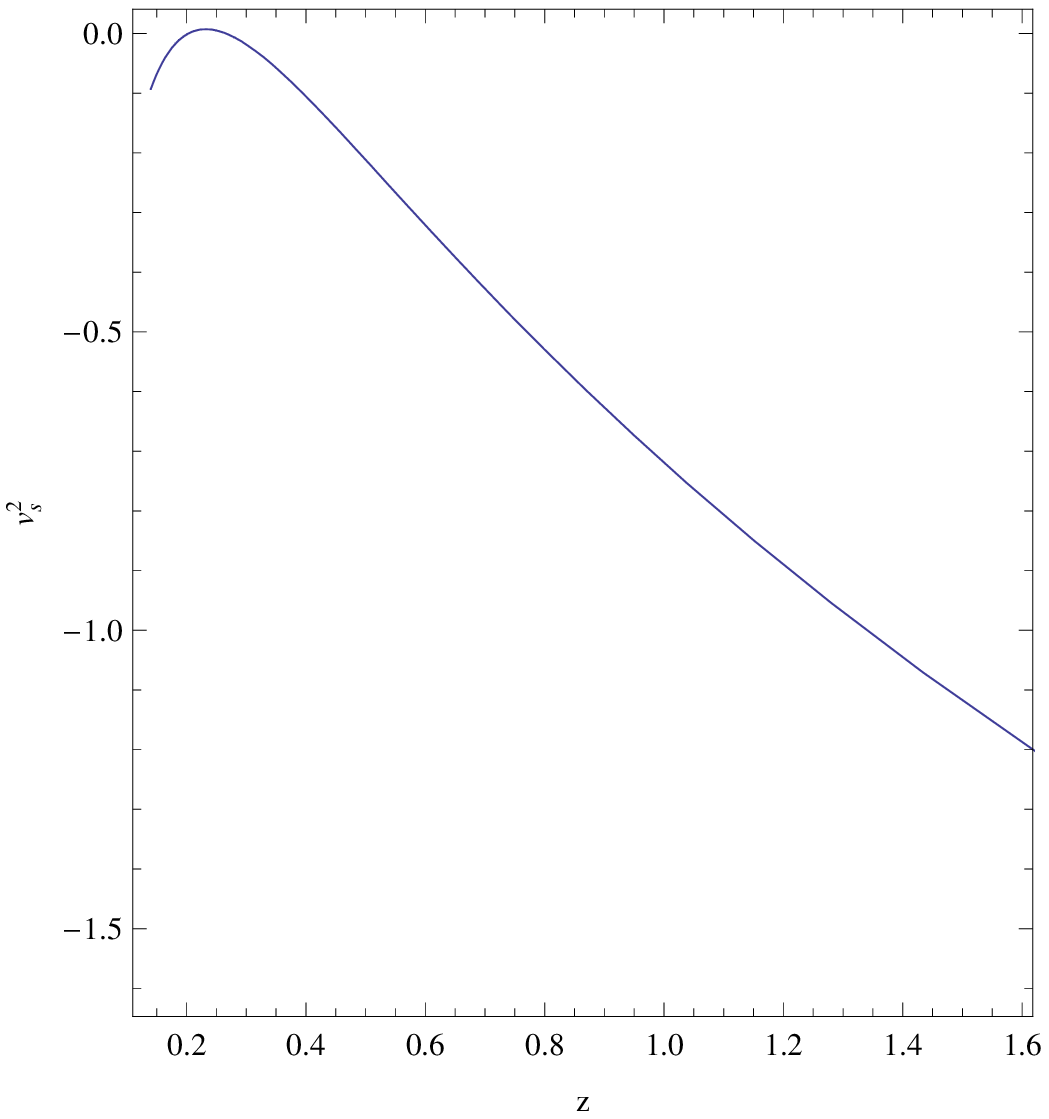}
\vspace{2 mm} ~~~~~~~~~~~~~~~~~~~~~~~Fig.31
~~~~~~~~~~~~~~~~~~~~~~~~~~~~~~~~~~~~~~~~~~~~~~~~~~~~~~~~~~~~~~~~~~~~~~~Fig.32

\textbf{Figs.25, 27, 26 and 28} represent the plots of $v_{s}^{2}$
for class I and class II scale factors in ECHDE $f(T)$ gravity
model in power-law correction. \textbf{Figs.29, 31, 30 and 32}
represent the plots of $v_{s}^{2}$ for class I and class II scale
factors in ECHDE $f(T)$ gravity model in logarithmic correction.
\end{figure}

\subsection{Investigation of stability of ECNADE in power-law and logarithmic
corrections:}

For ECNADE $f(T)$ model in power-law version there are also two
cases. For the first class (class I scale factor) we see from
\textbf{figures {\bf 33} and {\bf 35}} that $v_{s}^{2}>0$ for the
present and future epoch and for the second class (class II scale
factor) we see from \textbf{figures {\bf 34} and {\bf 36}} that
$v_{s}^{2}<0$ for the present and future epoch. So we can conclude
that ECNADE $f(T)$ model in power-law version implies a classical
stability for the first class and a classically instability of
second class of a given perturbation in
general relativity.\\

For ECNADE $f(T)$ model in logarithmic version there are also two
cases. For the first class (class I scale factor) we see from
figures {\bf 37} and {\bf 39} that $v_{s}^{2}>0$ for the present
and future epoch and for the second class (class II scale factor)
we see from figures {\bf 38} and {\bf 40} that $v_{s}^{2}>0$ . So
we can conclude that ECNADE $f(T)$ model in logarithmic version
implies a classical stability for the first
and second classes both.\\

\begin{figure}

\includegraphics[height=2.0in]{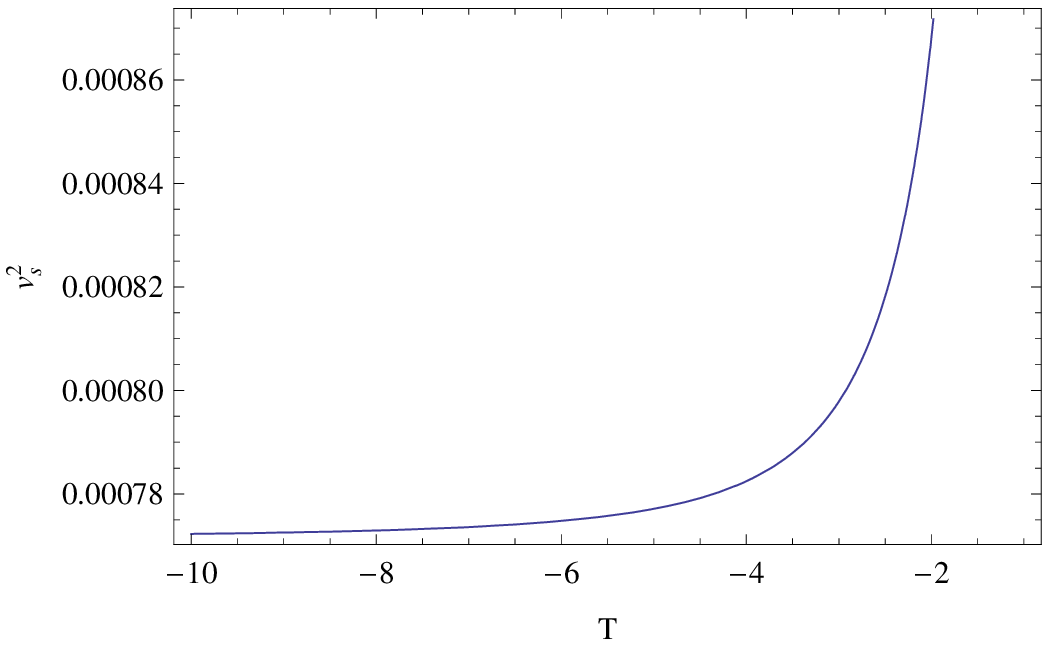}~~~~~
\includegraphics[height=2.0in]{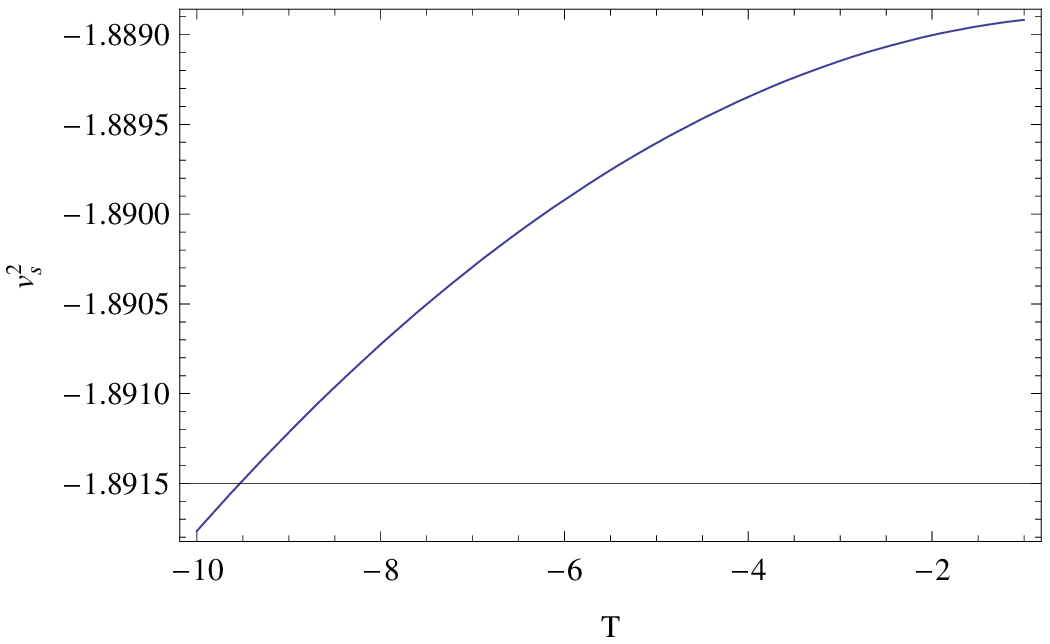}
\vspace{2 mm}
~~~~~~~~~~~~~~~~~~~~~~~Fig.33 ~~~~~~~~~~~~~~~~~~~~~~~~~~~~~~~~~~~~~~~~~~~~~~~~~~~~~~~~~~~~~~~~~~~~~~~Fig.34 \\
\vspace{1 mm}

\includegraphics[height=2.0in]{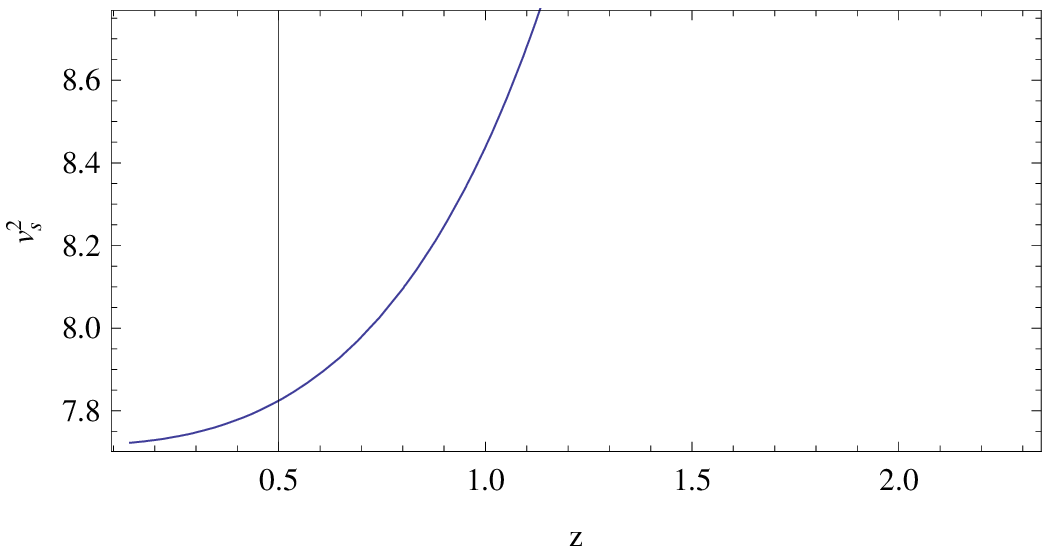}~~~~~
\includegraphics[height=2.0in]{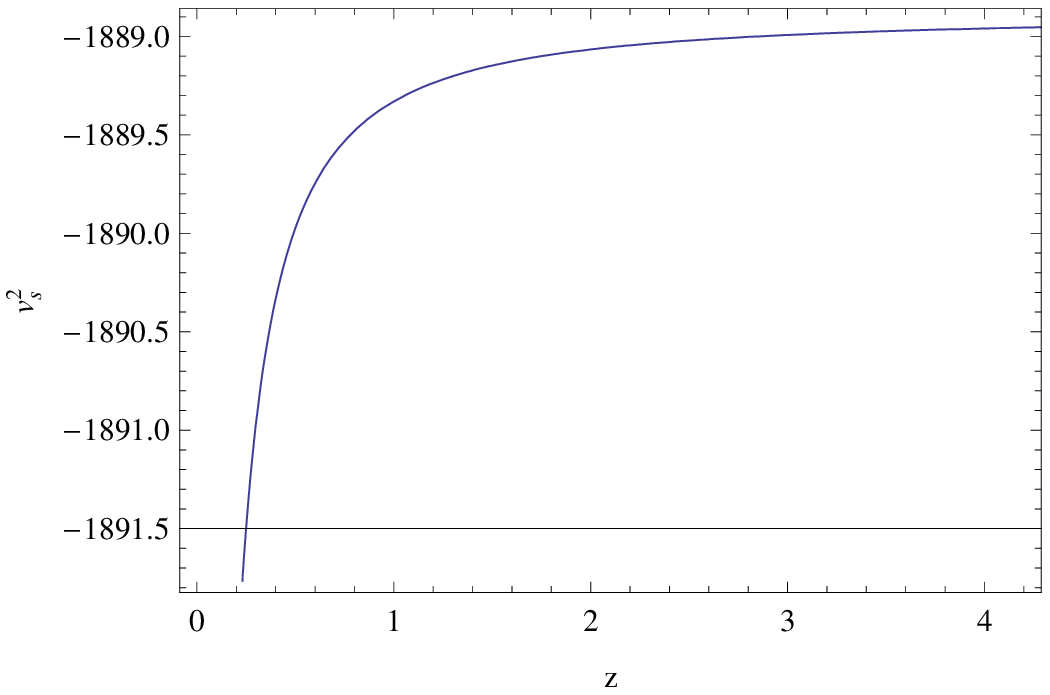}
\vspace{2 mm}
~~~~~~~~~~~~~~~~~~~~~~~Fig.35 ~~~~~~~~~~~~~~~~~~~~~~~~~~~~~~~~~~~~~~~~~~~~~~~~~~~~~~~~~~~~~~~~~~~~~~~Fig.36 \\
\vspace{1 mm}

\includegraphics[height=2.0in]{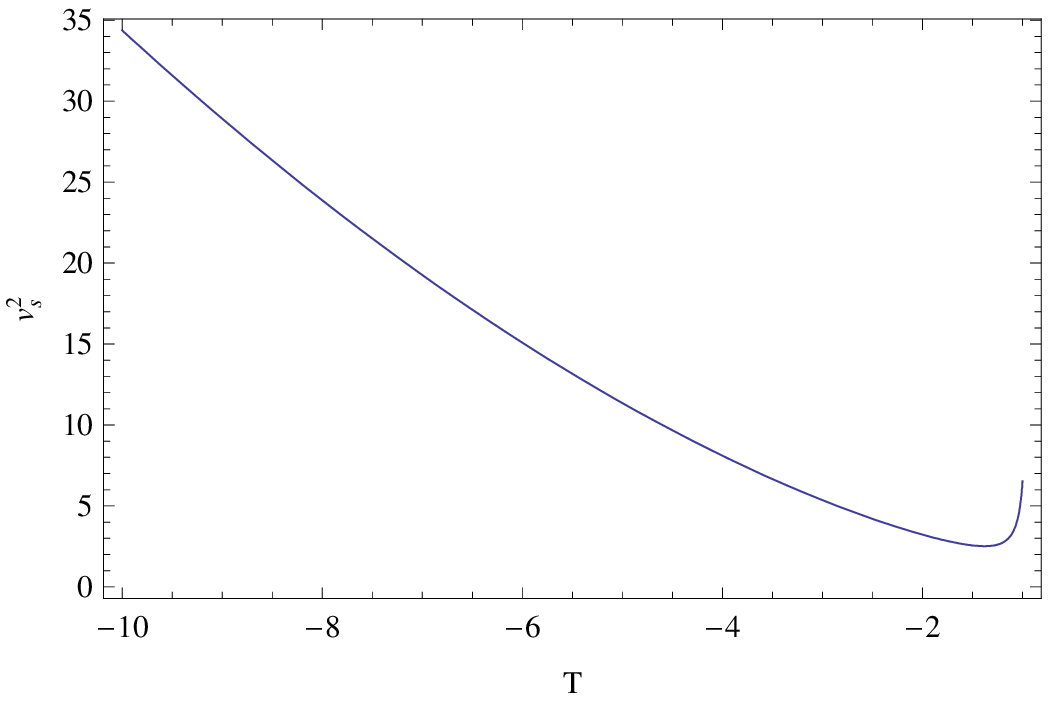}~~~~~
\includegraphics[height=2.0in]{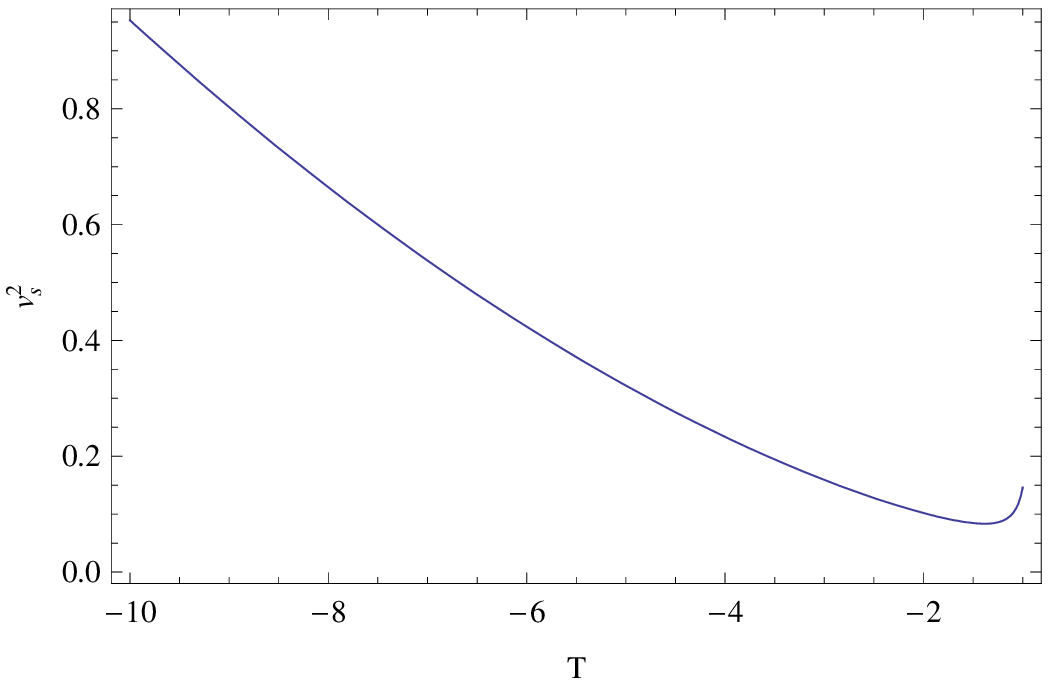}
\vspace{2 mm}
~~~~~~~~~~~~~~~~~~~~~~~Fig.37 ~~~~~~~~~~~~~~~~~~~~~~~~~~~~~~~~~~~~~~~~~~~~~~~~~~~~~~~~~~~~~~~~~~~~~~~Fig.38 \\
\vspace{1 mm}

\includegraphics[height=2.0in]{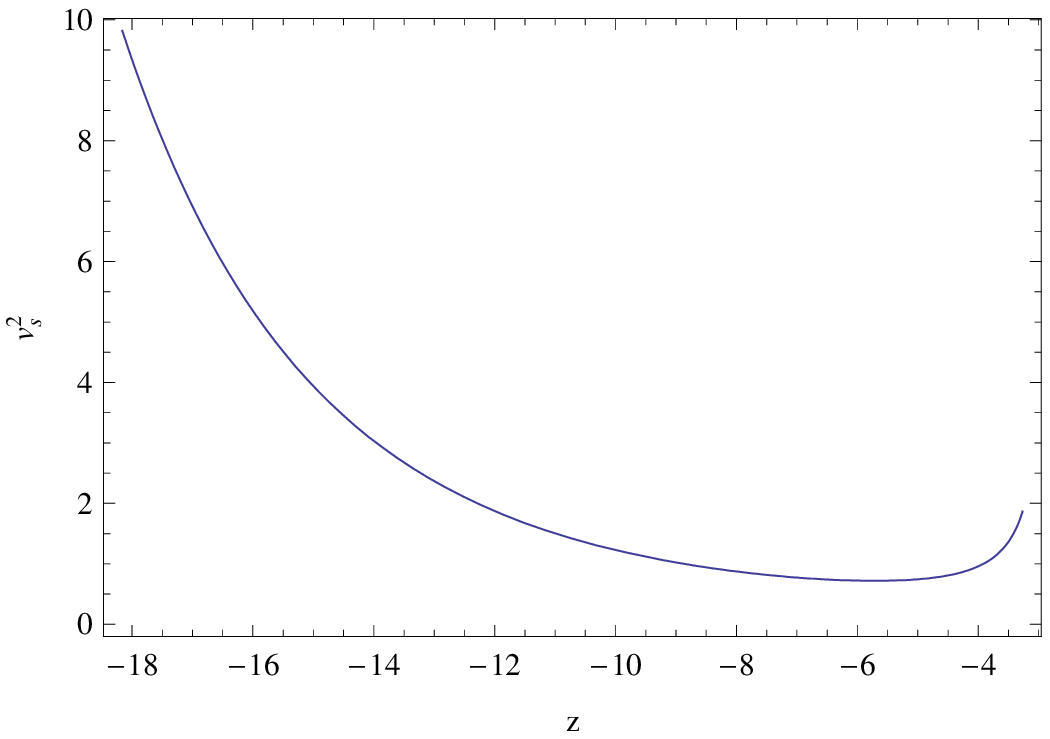}~~~~~
\includegraphics[height=2.0in]{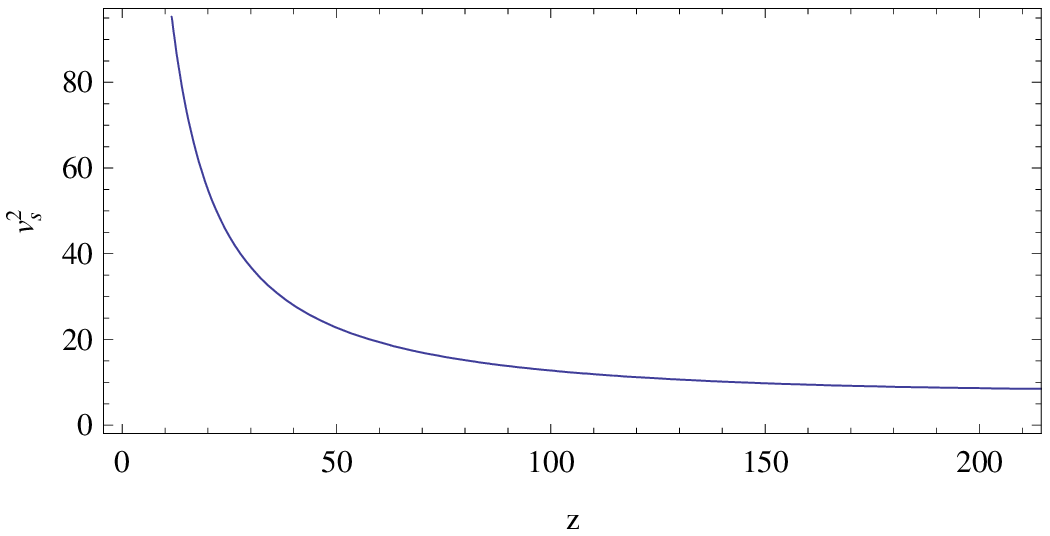}
\vspace{2 mm} ~~~~~~~~~~~~~~~~~~~~~~~Fig.39
~~~~~~~~~~~~~~~~~~~~~~~~~~~~~~~~~~~~~~~~~~~~~~~~~~~~~~~~~~~~~~~~~~~~~~~Fig.40

\textbf{Figs.33, 35, 34 and 36} represent the plots of $v_{s}^{2}$
for class I and class II scale factors in ECNADE $f(T)$ gravity
model in power-law correction. \textbf{Figs.37, 39, 38 and 40}
represent the plots of $v_{s}^{2}$ for class I and class II scale
factors in ECNADE $f(T)$ gravity model in logarithmic correction.
\end{figure}

\section{Concluding Remarks}

In this work, we have assumed the $f(T)$ modified gravity theory
in the background of flat FRW universe. We found the modified
Friedmann equations and then from the equations, we found the
effective energy density and pressure for $f(T)$ modified gravity
theory. Modified gravity gives a natural unification of the
early-time inflation and late-time acceleration. We have assumed
two types of power law forms of scale factor, the first class
(class I) has the future singularity and the second class (class
II) has the initial singularity. In the framework of $f(T)$
modified gravity model, four types of dark energy have been
considered, they are (i) entropy-corrected holographic dark energy
(ECHDE) in power-law version, (ii) entropy-corrected holographic
dark energy (ECHDE) in logarithmic version,(iii) entropy-corrected
new agegraphic dark energy (ECNADE) in power-law version and (iv)
entropy-corrected new agegraphic dark energy (ECNADE) in
logarithmic version, where, $R_{h}$ is assumed to be the future
event horizon and $\eta$ is assumed to be conformal time. Using
the two classes of scale factors, the unknown function $f(T)$ has
been found in term of $T$ for ECHDE and ECNADE models in power-law
and logarithmic versions. The corresponding equation of states
have also been generated. For the cases of ECHDE and ECNADE in
power-law and logarithmic versions the natures of $f(T)$ vs $T$
have been shown in figures {\bf 1}, {\bf 2}, {\bf 7}, {\bf 8},
{\bf 13}, {\bf 14}, {\bf 19}, {\bf 20}. For the cases of ECHDE in
power-law version (class I) and logarithmic version (class I and
II), ECNADE in power-law version (class I) and logarithmic version
(class I and II) the equation of state parameter $w_{\Lambda}$ has
been shown in \textbf{figures 3, 5, 9, 10, 11, 12, 15, 17, 21, 22,
23, 24} whereas in figures {\bf 3}, {\bf 5}; {\bf 21} , {\bf 23}
and {\bf 22} , {\bf 24} the EoS parameter is divergent at
$T=-1,-10$,$z=0,25$; $T=-4$, $z=-6$ and at $T=-2$, $z=13.5$. For
the cases of ECHDE in power-law version (class II) and ECNADE in
power-law version (class II) the equation of state parameter
$w_{\Lambda}$ has been shown in figures {\bf 4}, {\bf 6}, {\bf
16}, {\bf 18} and from the figures we have seen that these models
lie entirely in the phantom region. It should be mentioned that
Karami et al \cite{K} have investigated the $f(T)$ reconstructions
for HDE, NADE models and logarithmic versions of ECHDE, ECNADE
models only and for these models we got the similar expressions of
$f(T)$ but we have details studied the results graphically. To
examine the stability test for all the reconstructing models, we
have investigated the signs of the square of the velocity of
sound. For ECHDE model in power-law version , we have concluded
from figures {\bf 25} and {\bf 27}, that the corresponding model
is a classical stable for $T \le -2$, $z \ge 0.1$ and classically
unstable for $T \ge -2$, $z \le 0.1$ for the first class and from
figures {\bf 26} and {\bf 28} the corresponding model is a
classically unstable for second class of a given perturbation in
general relativity. For ECHDE model in logarithmic version , we
have concluded from figures {\bf 29}, {\bf 30}, {\bf 31}, {\bf 32}
that the corresponding models are unstable for class I and class
II both. On the other hand, for ECNADE model in power-law version
(class I), we have seen from figures {\bf 33}, {\bf 35} that the
corresponding model is stable and for ECNADE model in power-law
version (class II) we have seen from figures {\bf 34}, {\bf 36}
that the corresponding model is unstable. Again for ECNADE in
logarithmic version (class I and II), we have seen from figures
{\bf 37}, {\bf 38}, {\bf 39}, {\bf 40} the corresponding models
are stable. Thus we may conclude that our reconstructing ECHDE
model (class I), ECNADE model in power-law version (class I) and
logarithmic version (class I and II) are more realistic (and
classically stable) than the discussed
other models (classically unstable).\\

{\bf Acknowledgement:}\\

One of the author (UD) is thankful to IUCAA, Pune,
India for warm hospitality where part of the work was carried out.\\

\end{document}